\documentclass[apj,iop]{emulateapj}

\usepackage[colorlinks=true,citecolor=blue,linkcolor=cyan,breaklinks]{hyperref}
\usepackage{amssymb}
\usepackage{subfigure}
\usepackage{longtable}
\usepackage{amsmath}

\shorttitle{}
\shortauthors{Strait et al.}

\begin{document}

\title{RELICS: Properties of $\MakeLowercase{z}\geq5.5$ Galaxies Inferred from \textit{Spitzer} and \textit{Hubble} Imaging, Including A Candidate $\MakeLowercase{z}\sim6.8$ Strong [OIII] emitter}

\author{Victoria Strait\altaffilmark{1}}
\author{Maru{\v s}a Brada{\v c}\altaffilmark{1}}
\author{Dan Coe\altaffilmark{2}}
\author{Brian C. Lemaux\altaffilmark{1}}
\author{Adam C. Carnall\altaffilmark{4}}
\author{Larry Bradley\altaffilmark{2}}
\author{Debora Pelliccia\altaffilmark{3,23}}
\author{Keren Sharon}\altaffilmark{5}
\author{Adi Zitrin\altaffilmark{6}}
\author{Ana Acebron\altaffilmark{6}}
\author{Chloe Neufeld\altaffilmark{1}}
\author{Felipe Andrade-Santos\altaffilmark{7}}
\author{Roberto J. Avila\altaffilmark{2}}
\author{Brenda L. Frye\altaffilmark{8}}
\author{Guillaume Mahler\altaffilmark{5}}
\author{Mario Nonino\altaffilmark{9}}
\author{Sara Ogaz\altaffilmark{2}}
\author{Masamune Oguri\altaffilmark{10,11,12}}
\author{Masami Ouchi\altaffilmark{12,13}}
\author{Rachel Paterno-Mahler\altaffilmark{14}}
\author{Daniel P. Stark\altaffilmark{8}} 
\author{Ramesh Mainali\altaffilmark{24}} 
\author{Pascal A. Oesch\altaffilmark{16}} 
\author{Michele Trenti\altaffilmark{17,21}}
\author{Daniela Carrasco\altaffilmark{17}} 
\author{William A. Dawson\altaffilmark{18}}
\author{Christine Jones\altaffilmark{19}} 
\author{Keiichi Umetsu\altaffilmark{20}} 
\author{Benedetta Vulcani\altaffilmark{22}}
\affil{\altaffilmark{1}Physics and Astronomy Department, University of California, Davis, CA 95616, USA}
\affil{\altaffilmark{2}Space Telescope Science Institute, Baltimore, MD 21218, USA}
\affil{\altaffilmark{3}Physics and Astronomy Department, University of California, Santa Cruz, USA}
\affil{\altaffilmark{4}SUPA (Scottish Universities Physics Alliance), Institute for Astronomy, University of Edinburgh, Royal Observatory, Edinburgh EH9 3HJ, UK}
\affil{\altaffilmark{5}Department of Astronomy, University of Michigan, 1085 South University Ave, Ann Arbor, MI 48109, USA}
\affil{\altaffilmark{6}Department of Physics, Ben-Gurion University, Be’er-Sheva 84105, Israel}
\affil{\altaffilmark{7}Harvard-Smithsonian Center for Astrophysics, 60 Garden Street, Cambridge, MA 02138, USA}
\affil{\altaffilmark{8}Department of Astronomy, Steward Observatory, University of Arizona, 933 North Cherry Avenue, Tucson, AZ, 85721, USA}
\affil{\altaffilmark{9}INAF -- Osservatorio Astronomico di Trieste, via G. B. Tiepolo 11, I-34131
Trieste, Italy}
\affil{\altaffilmark{10}Research Center for the Early Universe, University of Tokyo, 7-3-1 
Hongo, Bunkyo-ku, Tokyo 113-0033, Japan}
\affil{\altaffilmark{11}Department of Physics, The University of Tokyo, 7-3-1 Hongo, Bunkyo-ku, Tokyo 113-0033, Japan}
\affil{\altaffilmark{12}Kavli Institute for the Physics and Mathematics of the Universe (Kavli IPMU, WPI), University of Tokyo, Kashiwa, Chiba 277-8583, Japan}
\affil{\altaffilmark{13}Institute for Cosmic Ray Research, The University of Tokyo, 5-1-5 Kashiwanoha, Kashiwa, Chiba 277-8582, Japan}
\affil{\altaffilmark{14}WM Keck Science Center, 925 N. Mills Avenue, Claremont, CA 91711}
\affil{\altaffilmark{15}Department of Physics and Astronomy, University of California, Riverside, CA 92521, USA}
\affil{\altaffilmark{16}Department of Astronomy, University of Geneva, Chemin des Maillettes 51, 1290 Versoix, Switzerland}
\affil{\altaffilmark{17}School of Physics, University of Melbourne, VIC 3010, Australia}
\affil{\altaffilmark{18}Lawrence Livermore National Laboratory, P.O. Box 808 L-210, Livermore, CA, 94551, USA}
\affil{\altaffilmark{19}Harvard-Smithsonian Center for Astrophysics, 60 Garden Street, Cambridge, MA 02138, USA}
\affil{\altaffilmark{20}Academia Sinica Institute of Astronomy and Astrophysics (ASIAA), No.~1, Section 4, Roosevelt Road, Taipei 10617, Taiwan}
\affil{\altaffilmark{21}ARC Centre of Excellence for All Sky Astrophysics in 3 Dimensions (ASTRO 3D), VIC 2010, Australia}

\affil{\altaffilmark{22}INAF-Osservatorio Astronomico di Padova, Vicolo Dell'osservatorio 5, 35122 Padova Italy}
\affil{\altaffilmark{23}UCO/Lick Observatories, University of California, Santa Cruz, 95065, USA}
\affil{\altaffilmark{24}Observational Cosmology Lab, NASA Goddard Space Flight Center, 8800 Greenbelt Rd., Greenbelt, MD 20771, USA}

\begin{abstract}
We present constraints on the physical properties (including stellar mass, age, and star formation rate) of 207 \mbox{$6\lesssim z \lesssim8$} galaxy candidates from the Reionization Lensing Cluster Survey (RELICS) and \emph{Spitzer}-RELICS surveys. We measure photometry using T-PHOT and perform spectral energy distribution fitting using EA$z$Y and BAGPIPES. Of the 207 candidates for which we could successfully measure (or place limits on) \textit{Spitzer} fluxes, 23 were demoted to likely $z<4$. Among the high-$z$ candidates, we find intrinsic stellar masses between $1\times10^6\rm{M_{\odot}}$ and $4\times10^9\rm{M_\odot}$, and rest-frame UV absolute magnitudes between $-22.6$ and $-14.5$ mag. While our sample is mostly comprised of $L_{UV}/L^*_{UV}<1$ galaxies, it extends to $L_{UV}/L^*_{UV}\sim2$. Our sample spans $\sim4$ orders of magnitude in stellar mass and star formation rates, and exhibits ages that range from maximally young to maximally old. We highlight 11 $z\geq6.5$ galaxies with detections in \emph{Spitzer}/IRAC imaging, several of which show evidence for some combination of evolved stellar populations, large contributions of nebular emission lines, and/or dust. Among these is PLCKG287+32-2013, one of the brightest $z\sim7$ candidates known (AB mag 24.9 at 1.6$\mu$m) with a \textit{Spitzer} 3.6$\mu$m flux excess suggesting strong [OIII] + H-$\beta$ emission ($\sim$1000\AA\ rest-frame equivalent width). We discuss the possible uses and limits of our sample and present a public catalog of \textit{Hubble} + \emph{Spitzer} photometry along with physical property estimates for all objects in the sample. Because of their apparent brightnesses, high redshifts, and variety of stellar populations, these objects are excellent targets for follow-up with the \textit{James Webb Space Telescope}.

\end{abstract}


\section{Introduction}
Properties of \mbox{$z \gtrsim 6$} galaxies are interesting not only for piecing together the role of galaxies in reionization, the period of time in the universe when energetic photons ionized neutral hydrogen in the intergalactic medium (IGM), but also as the key building blocks in galaxy formation models. Galaxies in this epoch often reveal characteristics rarely seen in local galaxies. Average stellar properties of galaxies up to $z\sim4-5$ have been reasonably well characterized; with access to a wealth of information from multiwavelength observations of characteristic galaxies, such as rest-frame optical and infrared (IR) data from Keck and Herchel for $z\sim2-3$, and the Atacama Large Millimeter Array (ALMA) for $z\sim1-5$, we have detailed accounts of basic physical properties such as star formation rate (SFR), stellar mass, and age, as well as metal enrichment and dust content for galaxies in this regime (e.g., \citealp{Sanders2020,Duncan2020,Fudamoto2020}). The picture becomes much less clear at higher redshifts, where due to intrinsic faintness of distant galaxies, increased absorption by the IGM, and the difficulty of obtaining red enough data to break degeneracies, we often have to rely on broadband imaging data and spectral energy distribution (SED) fitting for this information.

Broadband photometry and, in particular, \emph{Spitzer}/Infrared Array Camera (IRAC) fluxes play an important role in measuring physical properties of galaxies at $z\gtrsim6$ (see \citealp{Bradac2020} for a review). Because the rest-frame optical wavelengths are redshifted into the infrared in this regime, \emph{Spitzer}/IRAC $3.6\mu$m and $4.5\mu$m ([3.6] and [4.5] hereafter) observations are necessary for constraints of stellar mass and age until \emph{James Webb Space Telescope} (JWST) is functional. 

There have been a multitude of studies using \emph{Spitzer}/IRAC fluxes to probe the rest-frame optical wavelengths of high redshift galaxies. Notable examples include large surveys such as Hubble Frontier Fields (HFF, \citealp{Lotz2017,Merlin2016,Castellano2016b,DiCriscienzo2017,Santini2017,Shipley2018,Bradac2019}), Hubble Ultra Deep Field (HUDF, \citealp{Yan2005,Eyles2005,Labbe2010}), GOODS Re-ionization Era wide-Area Treasury from Spitzer (GREATS, \citealp{Stefanon2019}), Cluster Lensing and Supernova Survey with Hubble (CLASH, \citealp{Postman2012,Bouwens2014}), and Spitzer UltRa-Faint Survey (SURFSUP, \citealp{Bradac2014,Ryan2014,Huang2016a}). \emph{Spitzer}/IRAC [3.6] and [4.5] broadband imaging has been integral to facilitate and contextualize high-impact discoveries, such as evidence of evolved stellar populations at $z>8$ \citep{zheng12,Huang2016a,Hashimoto2018,Mawatari2020,Strait2020a}, discovery of the most distant spectroscopically confirmed galaxy, Gnz11 \citep{Oesch2016}, discovery of the highest-redshift Lyman-$\alpha$ detection \citep{Smit2015,Zitrin2015,Oesch2015,Roberts-Borsani2016,Jung2020}, measurement of nebular emission at $z\sim4$ \citep{Shim2011,Stark2013,Caputi2017,Bouwens2016ha,Faisst2016,Faisst2019}, and later measurement of nebular emission and stellar properties at $z>5$ \citep{Roberts-Borsani2016,Huang2016b,Stefanon2019,DeBarros2019,Laporte2014,Bridge2019,Roberts-Borsani2020}.

While the above surveys have established groundwork for observations of bright and faint galaxies at $z\gtrsim6$, open questions centered around characterizing high-redshift populations remain. Ionization field, dust content, metal enrichment, and ionizing photon production are some examples of still mostly unknown quantities for a ``normal" galaxy at $z\geq5.5$. The answers to these unknowns will require significant spectroscopic followup time with existing and future telescopes. 

In this work, we use \emph{Spitzer}/IRAC observations to measure physical properties and identify the most interesting galaxies for future spectroscopic follow-up. Unique to this work is the use of gravitational lensing of a large number of galaxy clusters to probe the most apparently bright but perhaps intrinsically fainter high redshift sources. The Reionization Cluster Lensing Survey (RELICS, PI Dan Coe) and companion survey \emph{Spitzer}-RELICS (S-RELICS), were designed to characterize the population of galaxies at these redshifts, and to attempt to find bright and rare galaxies at these epochs. To this end, these surveys image 41 massive clusters with \emph{Hubble Space Telescope (HST)} imaging data from RELICS for all 41 of these clusters, which was used to select 321 $z\geq5.5$ candidates \citep{Salmon2020}. Here we add the S-RELICS \emph{Spitzer}/IRAC imaging to further refine and characterize this sample. 

The structure of the paper is as follows: We describe HST and Spitzer imaging data and photometry in Section \ref{obsphot}, our SED modeling procedure and calculation of stellar properties in Section \ref{sedfit}, and lens models used for correction to relevant stellar properties in Section \ref{magmaps}. We present SED fitting results in Section \ref{results}, discuss possible future data in Section \ref{futuredata}, and we conclude in Section \ref{concls}. Throughout the paper, we will give magnitudes in the AB system \citep{Oke1974}, and we assume a $\Lambda$CDM cosmology with \mbox{$h = 0.7$}, \mbox{$\Omega_m = 0.3$}, and \mbox{$\Omega_{\Lambda} = 0.7$}. Equivalent widths are quoted in the rest-frame.

\section{Observations and Photometry}\label{obsphot}
All imaging data used for this analysis were obtained through a combination of RELICS and archival data. Each cluster was observed with \emph{HST} and \emph{Spitzer} and was reduced in a way that optimizes the search for high-$z$ galaxies. Here we briefly summarize the observing strategy of the survey, but in depth information about observations can be found on the RELICS website\footnote{https://relics.stsci.edu/} and in the RELICS overview paper \citep{Coe2019}. Images of selected \textit{HST} and bands and both \textit{Spitzer} channels for the $z\geq6.5$, IRAC-detected sample are shown in Figure \ref{fig:resids}.

\subsection{HST}
Each cluster was observed with two orbits of Wide Field Camera 3/ Infrared (WFC3/IR) imaging split among the F105W, F125W, F140W, and F160W filters, and three orbits of Advanced Camera for Surveys (ACS) split among F435W, F606W, and F814W (minus archival optical imaging), for a total of 188 \emph{HST} orbits. Several clusters had archival ACS and/or WFC3 imaging in other filters (F390W, F475W, F555W, F625W, F775W, F850LP, F110W) and some clusters which received additional data from a subsequent proposal. Most clusters reach approximate \textit{HST} depths of $\sim27$ mag (3-$\sigma$) in ACS bands and $\sim26$ mag (3-$\sigma$) in WFC3 bands.

In this paper, we use the catalogs based on a detection image comprised of the 0.06"/pixel weighted stack of all WFC3/IR imaging (to optimize the search for high-$z$ galaxies), described in \cite{Coe2019} and available on MAST\footnote{https://archive.stsci.edu/prepds/relics/}. 

\subsection{Spitzer Data and Photometry}
\emph{Spitzer}/Infrared Array Camera (IRAC) images for all clusters come from S-RELICS (\emph{Spitzer}-RELICS, PI Brada{\v c} \#12005, 13165, 13210, Director's Discretionary Time, PI Soifer \#12123), for a total of over 1000 hours of exposure time, and additional archival data. 13 clusters that had promising $z\sim8$ targets received deeper data via a follow-up proposal, to reach a total of 30 hours exposure time per band (3-$\sigma$ depth $\sim$26 mag) in each channel. All clusters reach a total of 5 hours of exposure time (3-$\sigma$ depth $\sim$24 mag) in each of IRAC channels 1 and 2 ([3.6] and [4.5]). A complete accounting of \emph{Spitzer} data can be found in Table \ref{tbl-1} and all shallow images are available on IRSA\footnote{https://irsa.ipac.caltech.edu/data/SPITZER/SRELICS/}. All raw data are available for download on the \textit{Spitzer} Heritage Archive (SHA\footnote{https://sha.ipac.caltech.edu/applications/Spitzer/SHA/}).

\begin{deluxetable*}{lllll}
\tabletypesize{\footnotesize}
\tablecaption{\label{tbl-1} 
Exposure Times and Programs of Spitzer data} 
\tablewidth{0pt}
\tablehead{
\colhead{Cluster} &  \colhead{$\rm{RA}^{\tablenotemark{1}}$} & \colhead{$\rm{Dec}^{\tablenotemark{1}}$} &\colhead{Exposure Time} &  \colhead{$\rm{Program}^{\tablenotemark{*}}$}}
PLCKG004-19 & 19:17:4.50 & $-$33:31:28.5  & 30, 30 hours &  13165, 12005, 12123 \\
SPT0615-57 & 06:15:54.2 & $-$57:46:57.9 & 30, 30 hours &  80012, 12005, 12123, 13210 \\
CL0152-13 & 01:52:42.9 & $-$13:57:31.0  & 30, 30 hours &  17, 20740, 50726, 70063, 12005, 12123, 14017 \\
ACT0102-49 & 01:02:53.1 & $-$49:14:52.8 & 30, 30 hours &  70149, 12123, 12005, 14017 \\
PLCKG287+32 & 11:50:50.8 & $-$28:04:52.2  & 30, 30 hours &  12123, 13165, 12005 \\
PLCKG308-20 & 15:18:49.9 & $-$81:30:33.6  & 30, 30 hours &  12123, 12005, 14017, 14253\\
MS1008-12 & 10:10:33.6 & $-$12:39:43.0 & 30, 30 hours &  12005, 12123, 14017 \\
RXS0603+42 & 06:03:12.2 & +42:15:24.7  & 30, 30 hours &  12005, 12123, 14017 \\
SMACS0723-73 & 07:23:19.5 & $-$73:27:15.6 & 30, 30 hours &  12123, 12005, 14017 \\
Abell1763 & 13:35:18.9 & +40:59:57.2 & 30, 30 hours &  12123, 13165, 12005 \\
MACS0553-33 & 05:53:23.1 & $-$33:42:29.9  & 30, 30 hours &  90218, 12005, 12123, 14281 \\
MACS0257-23 & 02:57:10.2 & $-$23:26:11.8  & 5, 5 hours &  60034 \\
RXC0600-20 & 06:00:09.8 & $-$20:08:08.9  &  5, 5 hours &  12005, 12123, 90218 \\
MACS0025-12 & 00:25:30.3 & $-$12:22:48.1  & 5, 5 hours &  60034 \\
Abell2163 & 16:15:48.3 & $-$06:07:36.7  & 9, 9 hours &  50096, 12005, 12123, 14242\\
Abell1758 & 13:32:39.0 & +50:33:41.8  & 6, 6 hours &  83, 60034 \\
RXC0018+16 & 00:18:32.6 & +16:26:08.4 & 5, 5 hours &  12005, 83, 12123 \\
Abell520 & 04:54:19.0 & +02:56:49.0  & 10, 10 hours &  12005, 12123 \\
MACS0308+26 & 03:08:55.7 & +26:45:36.8 & 5, 5 hours &  12005, 12123 \\
RXC0911+17 & 09:11:11.4 & +17:46:33.5  & 30, 30 hours &  60034, 14281 \\
Abells295 & 02:45:31.4 & $-$53:02:24.9 & 30, 30 hours &  70149, 12005, 12123, 14281 \\
Abell665 & 08:30:57.4 & +65:50:31.0  & 7, 5 hours &  12005, 12123, 14253 \\
Abell3192 & 03:58:53.1 & $-$29:55:44.8 & 5, 5 hours &  12123, 12005,  \\
PLCKG209+10 & 07:22:23.0 & +07:24:30.0 & 5, 5 hours &  12123, 12005 \\
Abell2537 & 23:08:22.2 & $-$02:11:32.4 & 5, 5 hours &  60034, 41011 \\
SPT0254-58 & 02:54:16.0 & $-$58:57:11.0 & 5, 5 hours &  12123, 12005 \\
RXC0142+44 & 01:42:55.2 & +44:38:04.3  & 5, 5 hours &  12123, 12005 \\
Abell1300 & 11:31:54.1 & $-$19:55:23.4 & 7, 5 hours &  12005, 12123, 14253, 14242 \\
MACS0159-08 & 01:59:49.4 & $-$08:50:00.0  & 7, 5 hours &  12005, 12123, 14253 \\
MACS0035-20 & 00:35:27.0 & $-$20:15:40.3  & 5, 5 hours &  12123, 12005 \\
WHL0137-08 & 01:37:25.0 & $-$08:27:25.0  & 7, 5 hours &  12123, 12005, 14253 \\
Abell697 & 08:42:58.9 & +36:21:51.1  & 5, 6 hours &  83, 60034, 14130, 14253 \\
PLCKG138-10 & 02:27:06.6 & +49:00:29.9 & 5, 5 hours &  12123, 12005 \\
PLCKG171-40 & 03:12:56.9 & +08:22:19.2 & 7, 5 hours &  12123, 12005, 14253 \\
RXC0032+18 & 00:32:11.0 & +18:07:49.0  & 5, 5 hours &  12123, 12005, 90218 \\
RXC0232-44 & 02:32:18.1 & $-$44:20:44.9 & 7, 5 hours &  12123, 12005, 14253 \\
RXC0949+17 & 09:49:50.9 & +17:07:15.3  & 5, 5 hours &  12123, 12005 \\
RXC1514-15 & 15:15:00.7 & $-$15:22:46.7 & 5, 5 hours &  12123, 12005, 14253 \\
RXC2211-03 & 22:11:45.9 & $-$03:49:44.7 & 6, 5 hours &  90218, 12005, 12123, 14253 \\
Abell2813 & 00:43:25.1 & $-$20:37:14.8  & 7, 5 hours &  60034, 14253\\
MACS0417-11 & 04:17:33.7 & $-$11:54:22.6  & 5, 5 hours &  12123, 12005, 90218 \\
\tablenotetext{1}{RAs and Decs correspond to cluster centers.}
\tablenotetext{*}{Program IDs included in our reduction. ID \#12005, 13165, 14281, 13165, 13210, 14017: PI Maru{\v s}a Brada{\v c}. \#12123: PI Tom Soifer, \#60034, 90218 PI Eiichi Egami,  \#14253 PI Mauro Stefanon, \#14242  PI Andra Stroe, \#50096 PI Paul Martini, \#70149 PI Felipe Menanteau, \#83  PI George Rieke, \#14130 PI Rychard J Bouwens, \#17  PI Giovanni Fazio, \#20740, 50726, 70063 PI Bradford P Holden, \#80012 PI Mark Brodwin
}

\end{deluxetable*}

\begin{figure*}[h!!!]
    \vspace{-0.2cm}
    \begin{subfigure}
        \centering
        \includegraphics[width=18cm]{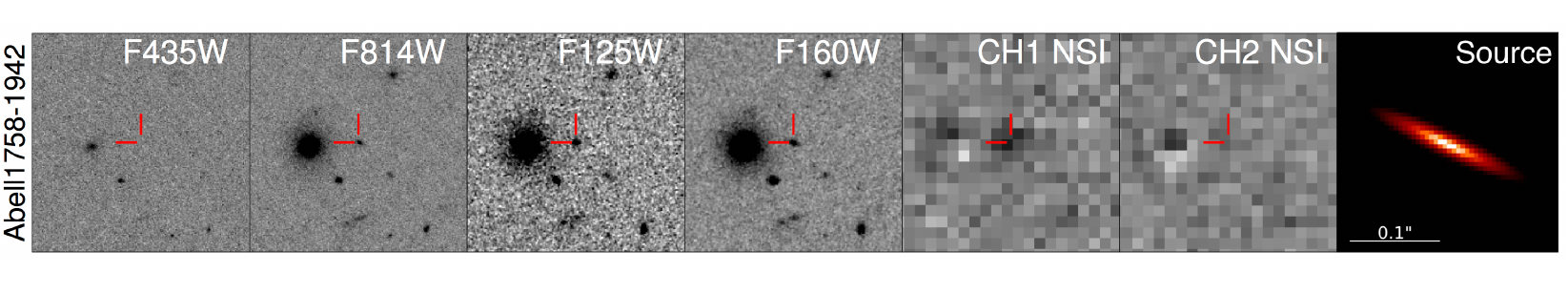}
    \end{subfigure}
    \vspace{-0.3cm}
    \begin{subfigure}
        \centering
        \includegraphics[width=18cm]{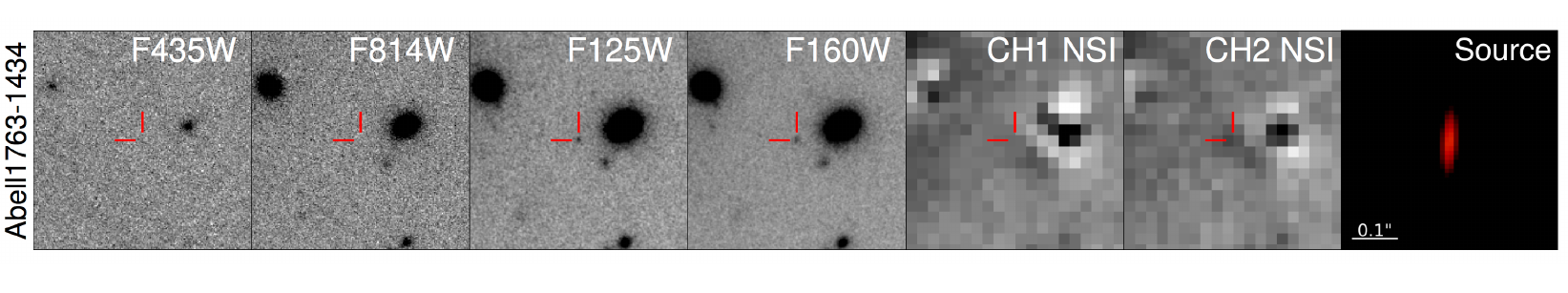}
    \end{subfigure}
    \vspace{-0.3cm}
    \begin{subfigure}
        \centering
        \includegraphics[width=18cm]{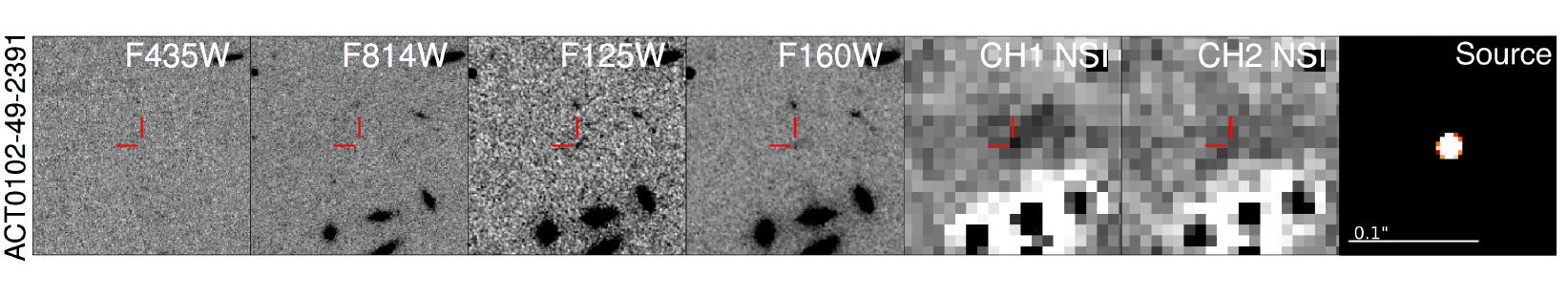}
    \end{subfigure}
    \vspace{-0.3cm}
    \begin{subfigure}
        \centering
        \includegraphics[width=18cm]{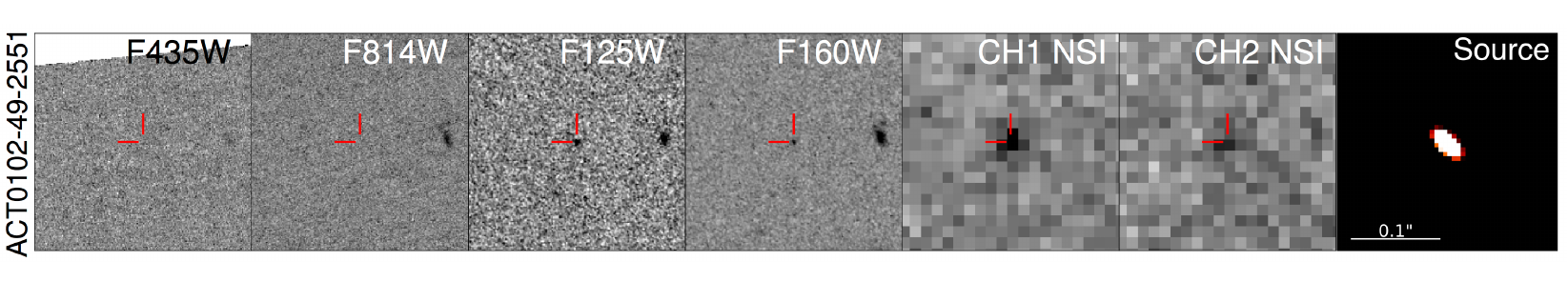}
    \end{subfigure}
    \vspace{-0.3cm}
    \begin{subfigure}
        \centering
        \includegraphics[width=18cm]{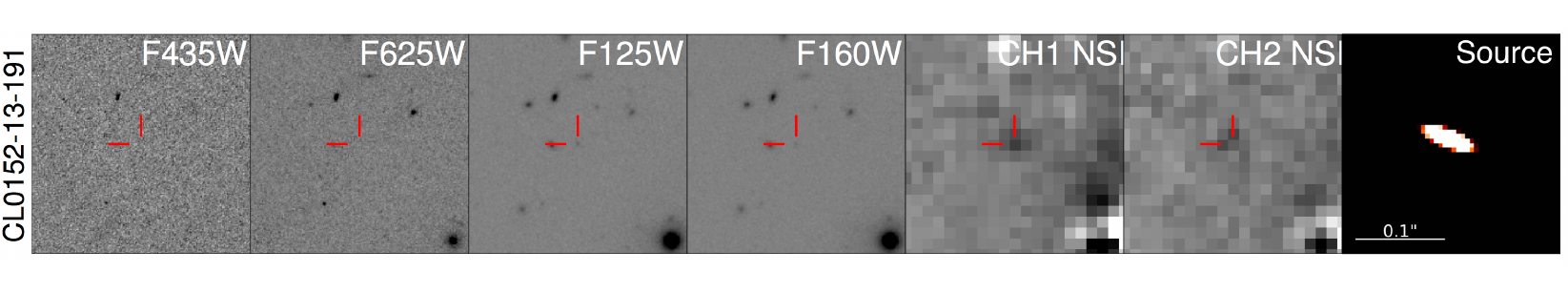}
    \end{subfigure}
    \vspace{-0.3cm}
    \begin{subfigure}
        \centering
        \includegraphics[width=18cm]{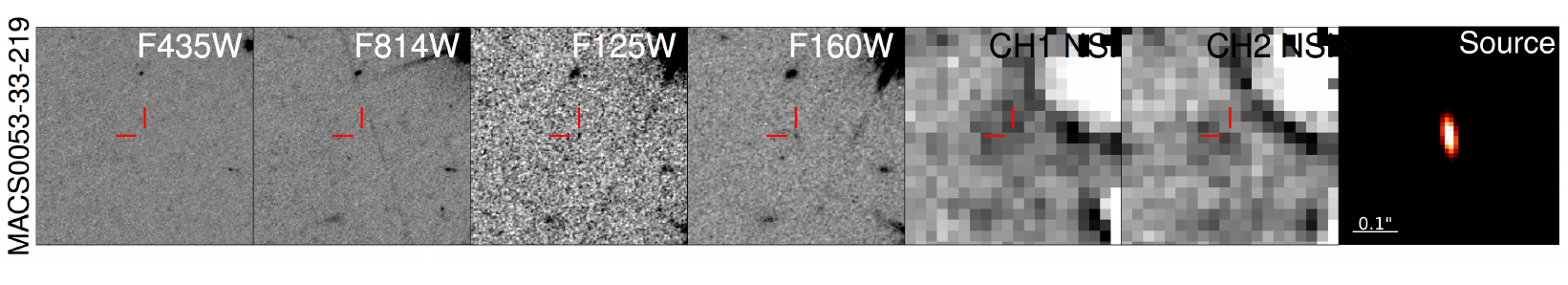}
    \end{subfigure}
    \vspace{-0.3cm}
    \begin{subfigure}
        \centering
        \includegraphics[width=18cm]{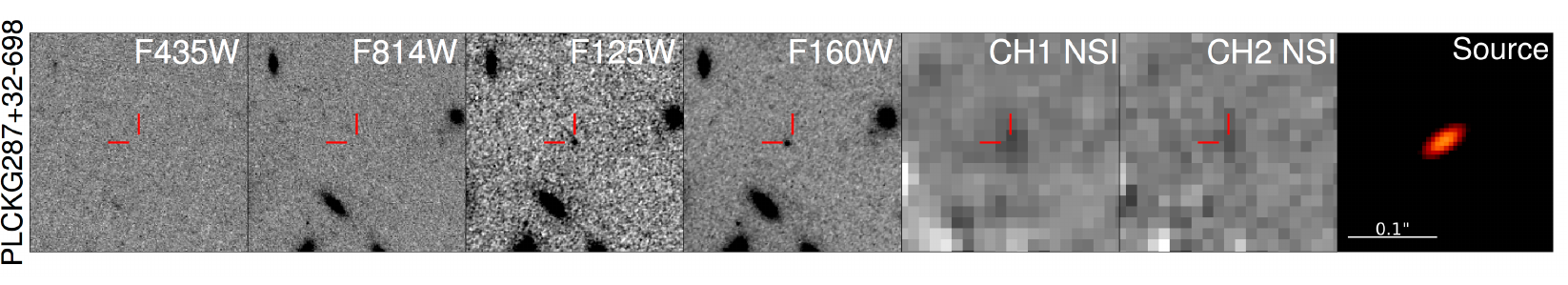}
    \end{subfigure}
\end{figure*}
\begin{figure*}
    \begin{subfigure}
        \centering
        \includegraphics[width=18cm]{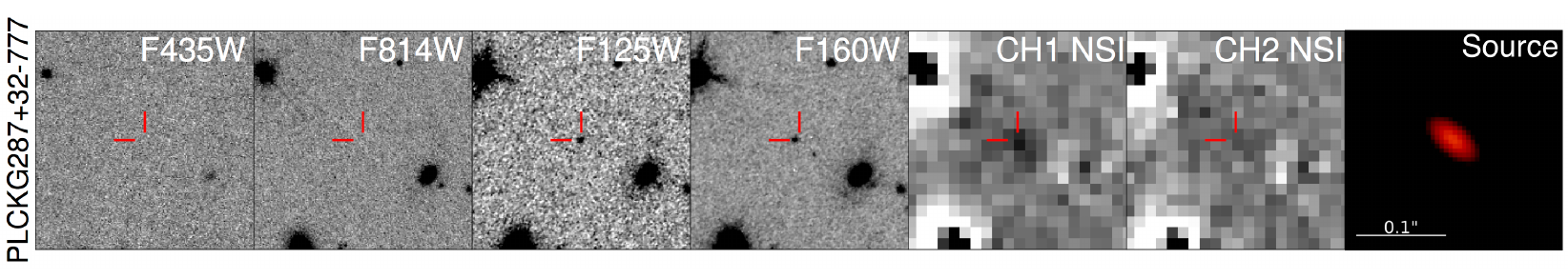}
    \end{subfigure}
    \vspace{-0.3cm}
    \begin{subfigure}
        \centering
        \includegraphics[width=18cm]{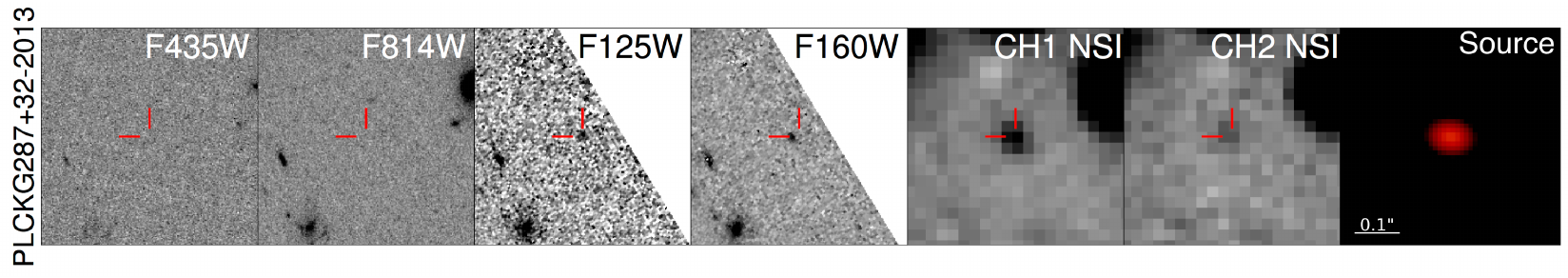}
    \end{subfigure}
    \vspace{-0.3cm}
    \begin{subfigure}
        \centering
        \includegraphics[width=18cm]{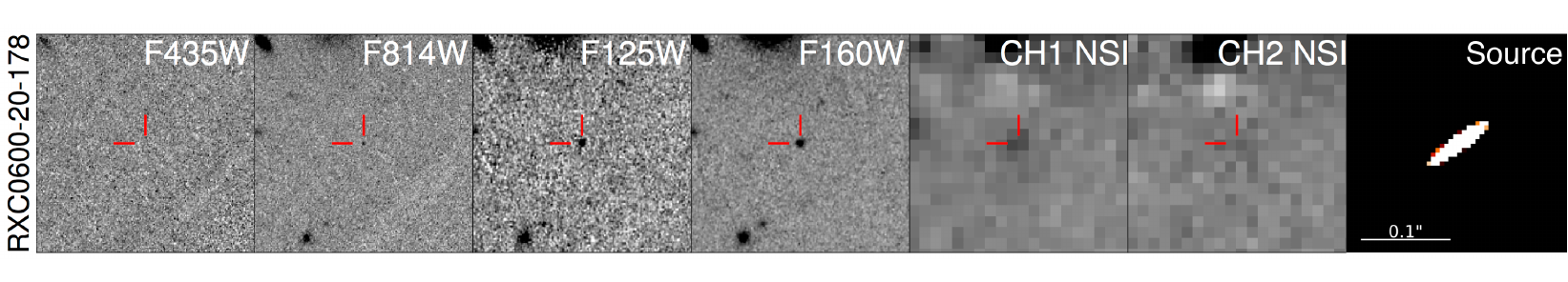}
    \end{subfigure}
    \vspace{-0.3cm}    
    \begin{subfigure}
        \centering
        \includegraphics[width=18cm]{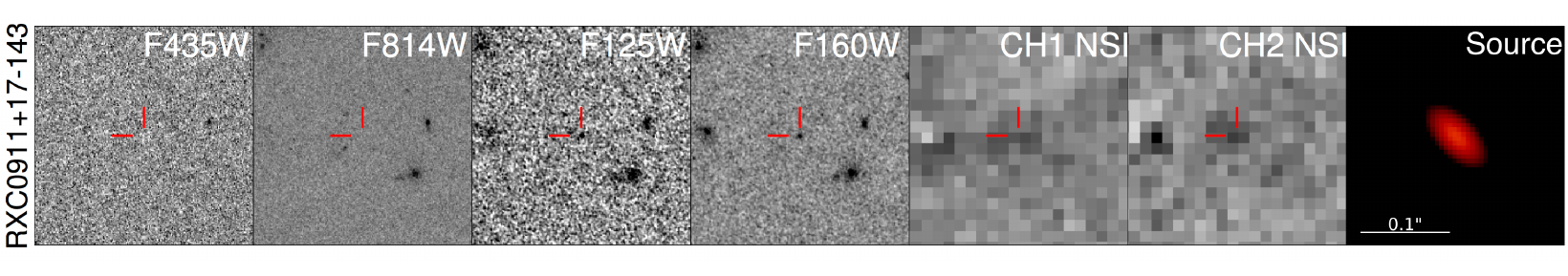}
    \end{subfigure}
    \caption{
    \emph{HST} images and \emph{Spitzer}/IRAC neighbor-subtracted images (NSIs) for each of the 11 galaxies that have both a best fit redshift of $z \geq 6.5$ and at least one \emph{Spitzer}/IRAC detection. 
    From left to right, F435W, F814W or F850LP (ACS, where $z > 7$ galaxies should not be detected), F125W, F160W (WFC3/IR), and Ch1, Ch2 of \emph{Spitzer}. The cutouts are 12"x12", and the object is centered near the red tick marks in all but the right-most panel. The rightmost panel is the resulting delensed source in the source plane of the galaxy with a 0.1" size bar for comparison. }
    \label{fig:resids}
\end{figure*}

\subsubsection{Reduction}
To reduce and mosaic \textit{Spitzer} images, we closely follow the process described by \cite{Bradac2014} for the SURFSUP survey, which also consists of IRAC images used for a search of high-$z$  galaxies. Briefly, we begin downloading the corrected-basic calibrated data (cBCD) from the SHA. The cBCDs have been processed by the IRAC pipeline to remove instrumental artifacts, and to calibrate into physical units (MJy $\rm{sr}^{-1}$). We apply additional mitigation measures using custom scripts; specifically, we correct for the warm-mission column pulldown (using bandcor\_warm.c by M. Ashby) and muxstriping (using automuxstripe.pro by J. Surace) that often occurs when bright sources are present (these scripts are located on the IRAC Cookbook website\footnote{https://irsa.ipac.caltech.edu/data/SPITZER/docs/\\dataanalysistools/cookbook/}). To create the mosaic images we use the MOsaicker and Point source EXtractor (MOPEX) command-line tools and largely follow the process described in the IRAC Cookbook for the COSMOS medium-deep data.
The corrected frames are background-matched using the overlap.pl routine from MOPEX and then drizzle-combined into a mosaic using the mosaic.pl routine. The final mosaics have a pixel scale of 0.6"$\rm{pix}^{-1}$ and a pixel fraction of 0.1. As a last step, we use the \texttt{Tweakreg} routine from \texttt{DrizzlePac}, which compares the positions of bright objects in \textit{Spitzer} and \textit{HST} images, to correct for any shifts in relative astrometry. 


\subsubsection{Flux Extraction and Error Analysis}\label{fluxextract}
Intracluster light (ICL) subtraction, background subtraction, and flux extraction is done using T-PHOT \citep{merl15}, designed to perform PSF-matched, prior-based, multi-wavelength photometry on low-resolution imaging as described by \cite{merl15,Merlin2016}. This is done by convolving cutouts from a high resolution image (in the case of this work, \emph{HST}/WFC3 F160W) using a low resolution PSF transformation kernel that matches the high-resolution image to the low-resolution (in our case, [3.6] and [4.5]) image. T-PHOT then fits a template to each source detected with HST and convolves the template with a PSF transformation kernel to match the resolution to that of the IRAC image. T-PHOT then solves for the solution where the model image created from the convolved \emph{HST} image best matches the pixel values in the real IRAC image and outputs fluxes for each template. We use the F160W image and WFC3/IR total weighted segmentation map as the priors for T-PHOT. Because T-PHOT takes a template-fitting approach where all templates are solved for simultaneously, an approach designed for blank fields which have a zero mean background, we fit each high-$z$ candidate separately, running T-PHOT on a small FOV centered on the candidate, to account for the changing background and ICL in a cluster field. Within the 12" ($\sim70$ kpc for cluster redshift of $z=0.4$) FOV on which we run T-PHOT, the background shows minimal variation: in most cases, where the object is far from the cluster center, we see no variation in background (i.e., the variation is centered on zero and random), while this can reach as high as $\sim20$\% where ICL is denser near the cluster center. For these reasons, and due to the fact that most of our galaxies are far from the cluster center where ICL is not dense, further ICL modeling and subtraction are not necessary in addition to subtracting background.

While T-PHOT is useful for predicting fluxes of objects with potential blending, it does not account for this blending in its output uncertainties. However, it calculates a covariance matrix which includes the covariance between each object and every other object in the image. This can be used to calculate a maximum covariance index ($\rm{R_{[3.6],[4.5]}}$), which is the ratio of the covariance of an object with its closest or brightest neighbor and its own variance. If this value is higher than 1, the object is severely blended with a neighbor and caution should be taken interpreting fluxes. We report the R$_{\rm{[3.6], [4.5]}}$ values with the fluxes in Table \ref{tbl-2} and include them in the catalog for all sources in the sample. 

In practice, T-PHOT does not work as well near bright sources and the cluster center, creating artifacts in the residual (see several objects in Figure \ref{fig:resids}). For this reason, we only include fluxes in this work for which we believe we are able to reliably extract \emph{Spitzer} fluxes, meaning the residual on top of the high-$z$ candidate is not a residual artifact due to a bright nearby source. To ensure T-PHOT is not underestimating flux uncertainties, we calculate our own statistical uncertainties for each source. In the residual image, we measure background levels 100 times within 3" of each high-$z$ candidate, avoiding artifacts from neighboring objects. The mean should fall close to zero, and we take the standard deviation as the error. In most cases, this error is smaller than the error reported by T-PHOT, however in some cases it is larger. In those cases, we use the larger uncertainty.

The PSF and convolution kernel used to convolve prior HST images to the resolution of IRAC images are important for this process. We create a PSF by stacking point sources, identified with a Source Extractor \citep{bertin1996} run with the following parameters: \mbox{$\texttt{DEBLEND\_MINCONT}=10^{-7}$}, \mbox{$\texttt{MINAREA}=9$}, \mbox{$\texttt{DETECT\_THRESH}=2$}, and \mbox{$\texttt{ANALYSIS\_THRESH}=2$}. We select point sources using the stellar locus in \texttt{$\texttt{FLUX\_RADIUS}$} vs. \texttt{$\texttt{MAG\_AUTO}$} space for every object in the IRAC image (not just in the \emph{HST} FOV). After the point sources are selected, we further require that their axis ratio is $>0.9$ and that they are sufficiently separated from neighbors to have a secure centroid. We recompute the PSF centroids by fitting a Gaussian profile to the inner profile ($r<4$ pixels), and align the point sources using sinc interpolation. At each phase we subtract the local background and normalize the flux of the point source to one. We sigma-clip average the masked, registered, normalized point sources and do one further background correction to ensure the convolutions with T-PHOT are flux conserving. Our stacked PSFs are consistent with the IRAC handbook\footnote{https://irsa.ipac.caltech.edu/data/SPITZER/docs/irac/\\iracinstrumenthandbook/}, and each of our clusters' PSFs contains at least 40 point sources per IRAC channel. In practice, the PSF convolution kernel needs to be ``sharpened" to produce cleaner residuals. For each individual high-$z$ candidate, we experiment with increasing and decreasing the weight of the core of the PSF to produce the cleanest residual. Most of the time, this means decreasing the weight of the core by a factor of 0.8-0.9.

\begin{deluxetable*}{lcc}
\tabletypesize{\footnotesize}
\tablecaption{\label{tbl-4} Properties of SED Fitting Methods} 
\tablewidth{0pt}
\tablehead{
\colhead{} &\colhead{Method A} &  \colhead{Method B}
}
\startdata
Software & EA$z$Y & BAGPIPES \\
Redshift & 0 -- 12 & 4 -- 11 \\
Formation age & 10Myr -- age of universe &1 Myr -- age of universe \\
Metallicity & $0.2Z_{\odot}$ & 0.005 -- 5 $Z_\odot$ \\
Ionization log(U) & not known \textit{a priori} &-4 -- -1 \\
Dust & SMC, E(B-V) = 0 -- 1 & Calzetti, $\rm{A_V}$ = 0 -- $3\rm{mag}^\tablenotemark{1}$ \\
Templates & BC03+nebular emission & BPASS+CLOUDY nebular emission\\
SFH & constant & variable; see Eqn. 1-2

\tablenotetext{1}{2 times more dust surrounding HII regions for their first 10 Myr}

\end{deluxetable*}

\section{Estimating Galaxy Properties}\label{sedfit}
\begin{figure*}[h!!!]
    \begin{subfigure}
        \centering
        \includegraphics[width=6.4cm]{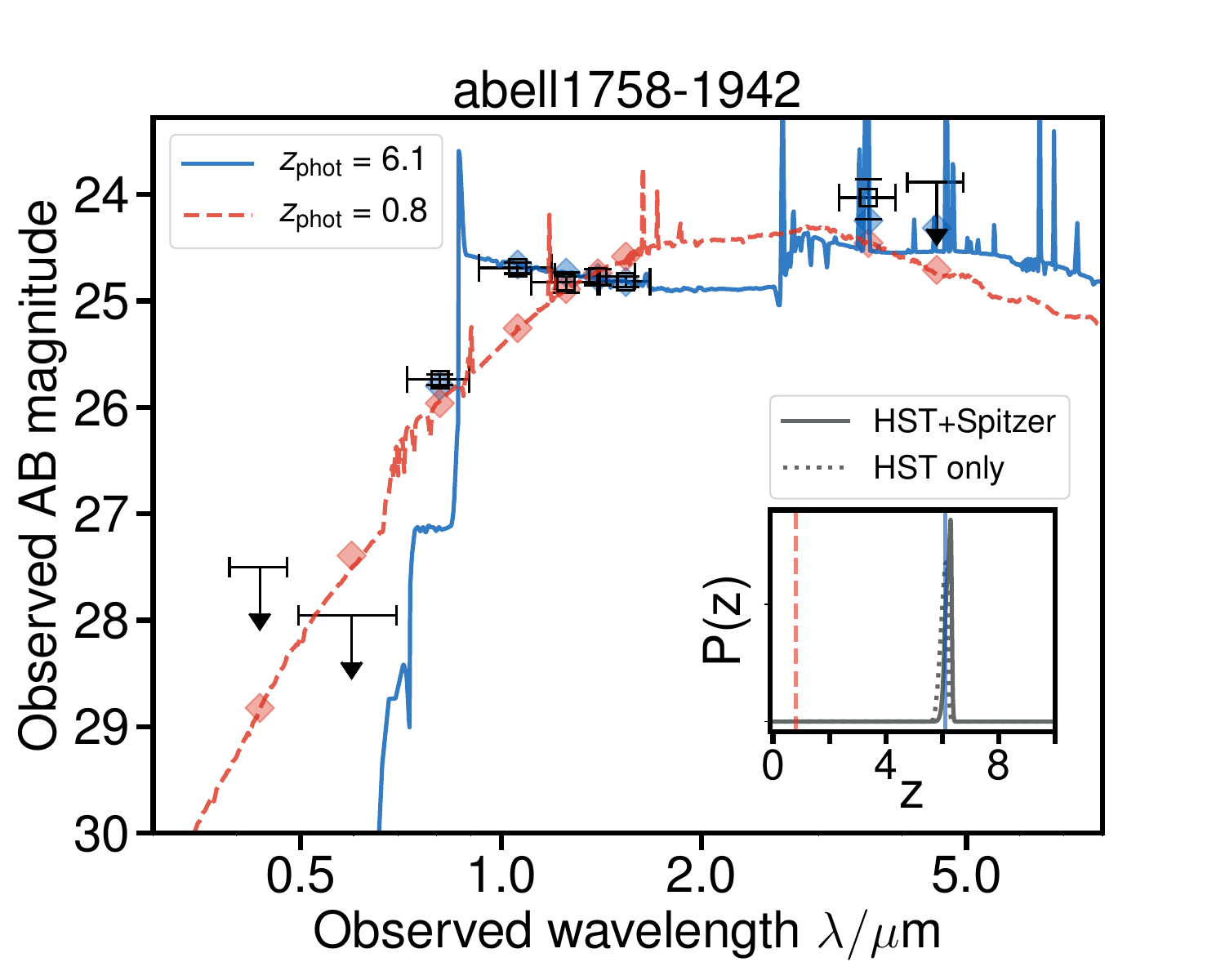}
    \end{subfigure}
    \hspace{-0.7cm}
    \begin{subfigure}
        \centering
        \includegraphics[width=6.4cm]{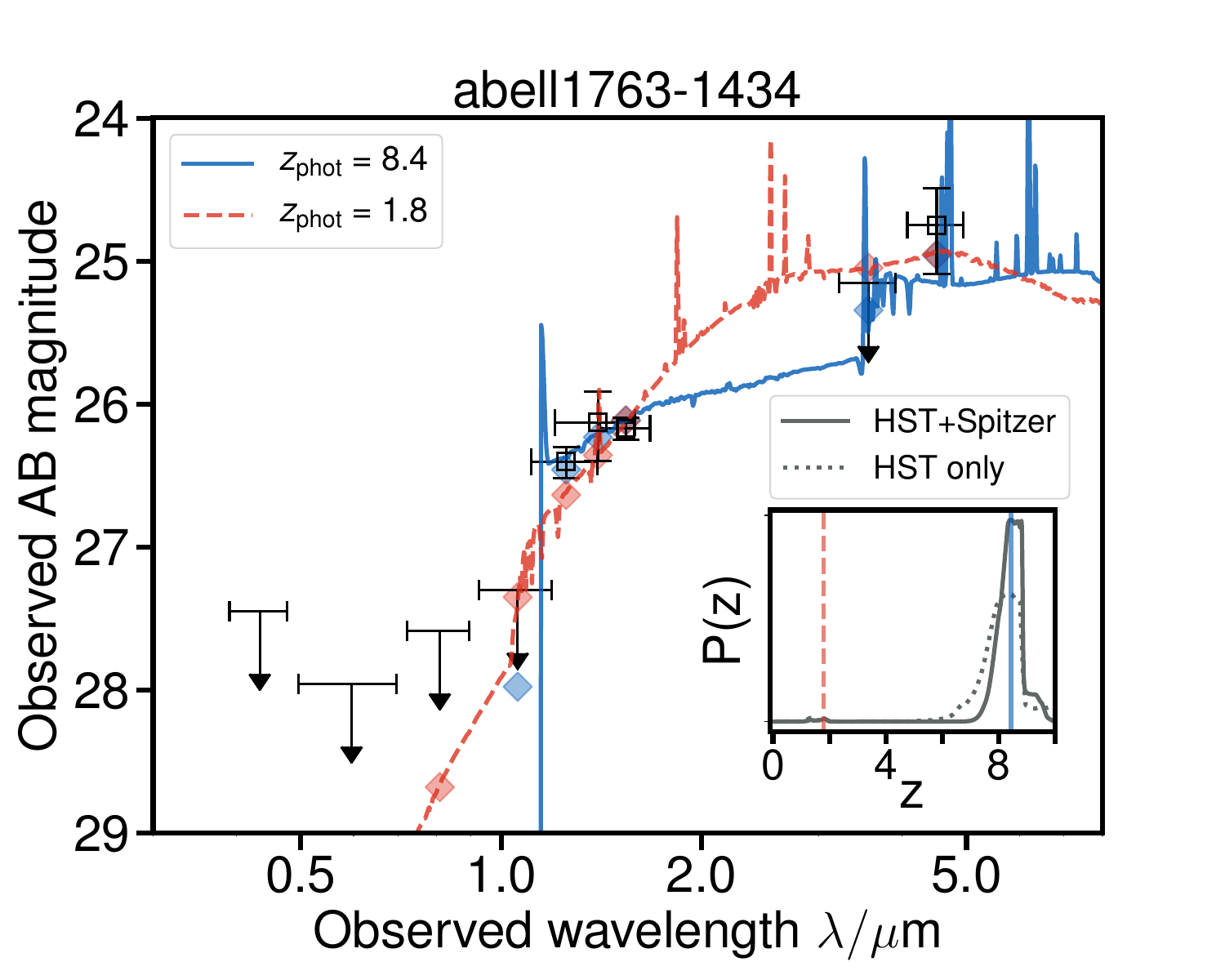}
    \end{subfigure}
    \hspace{-0.7cm}
    \begin{subfigure}
        \centering
        \includegraphics[width=6.4cm]{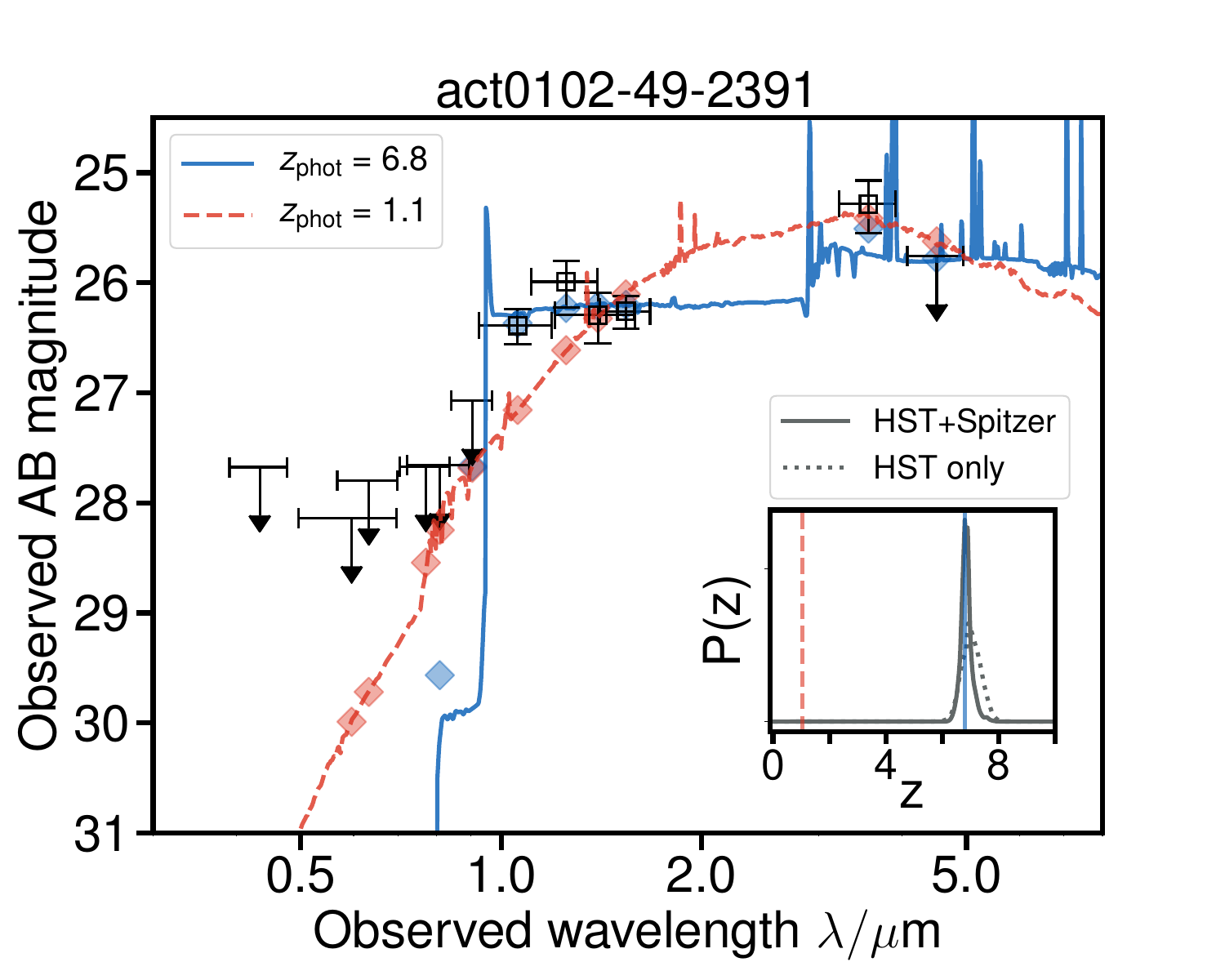}
    \end{subfigure}
    \begin{subfigure}
        \centering
        \includegraphics[width=6.4cm]{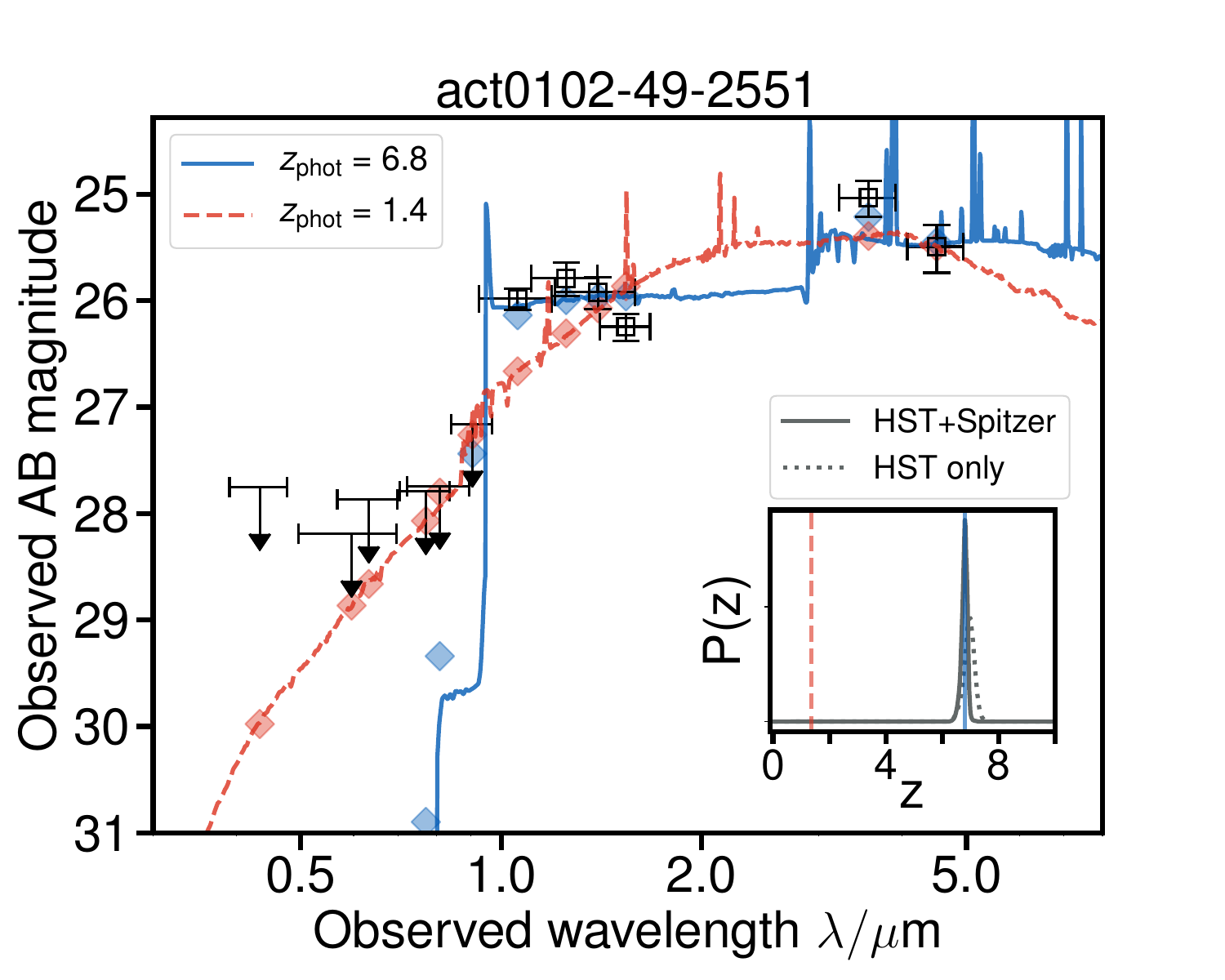}
    \end{subfigure}
    \hspace{-0.7cm}
    \begin{subfigure}
        \centering
        \includegraphics[width=6.4cm]{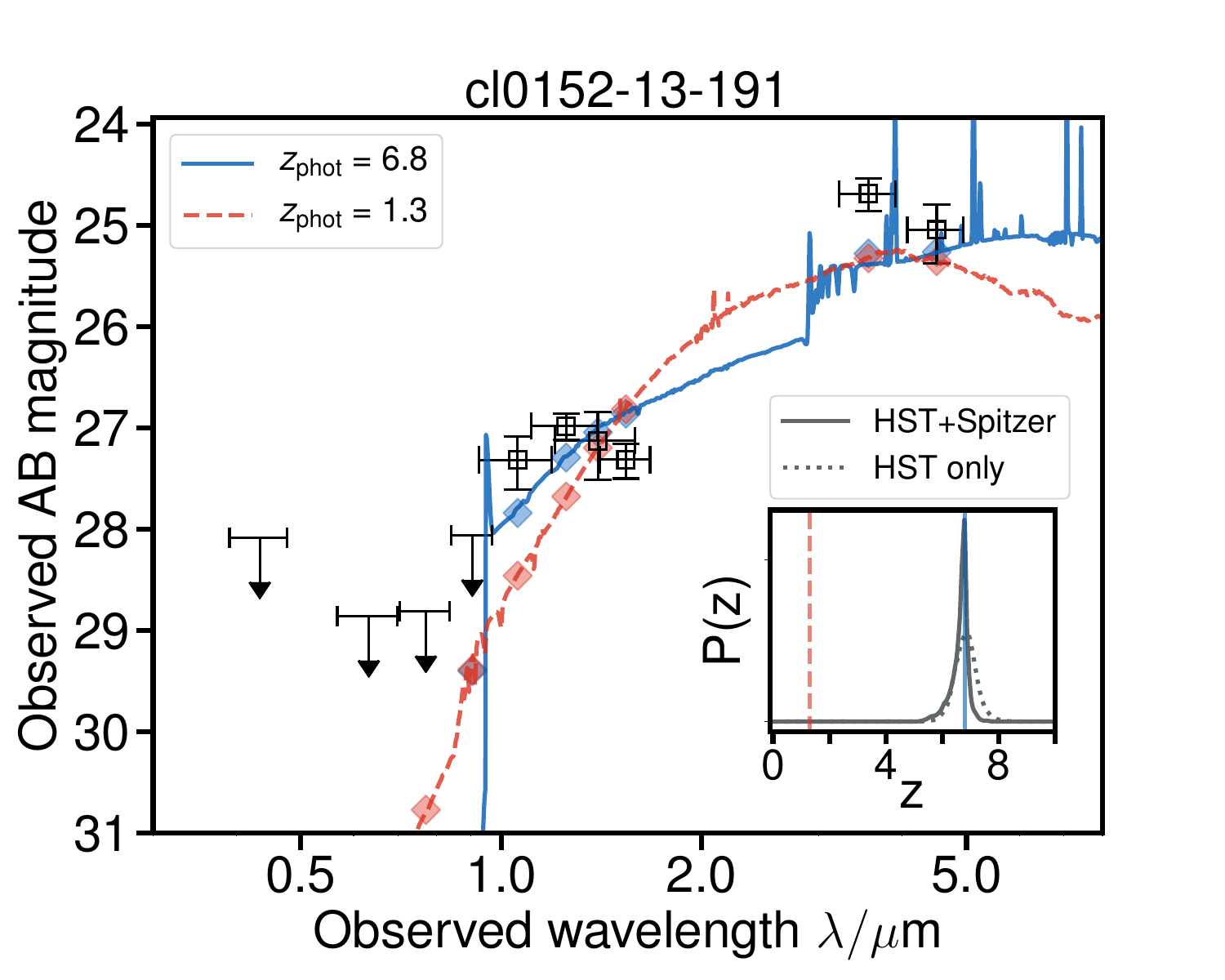}
    \end{subfigure}
    \hspace{-0.7cm}
    \begin{subfigure}
        \centering
        \includegraphics[width=6.4cm]{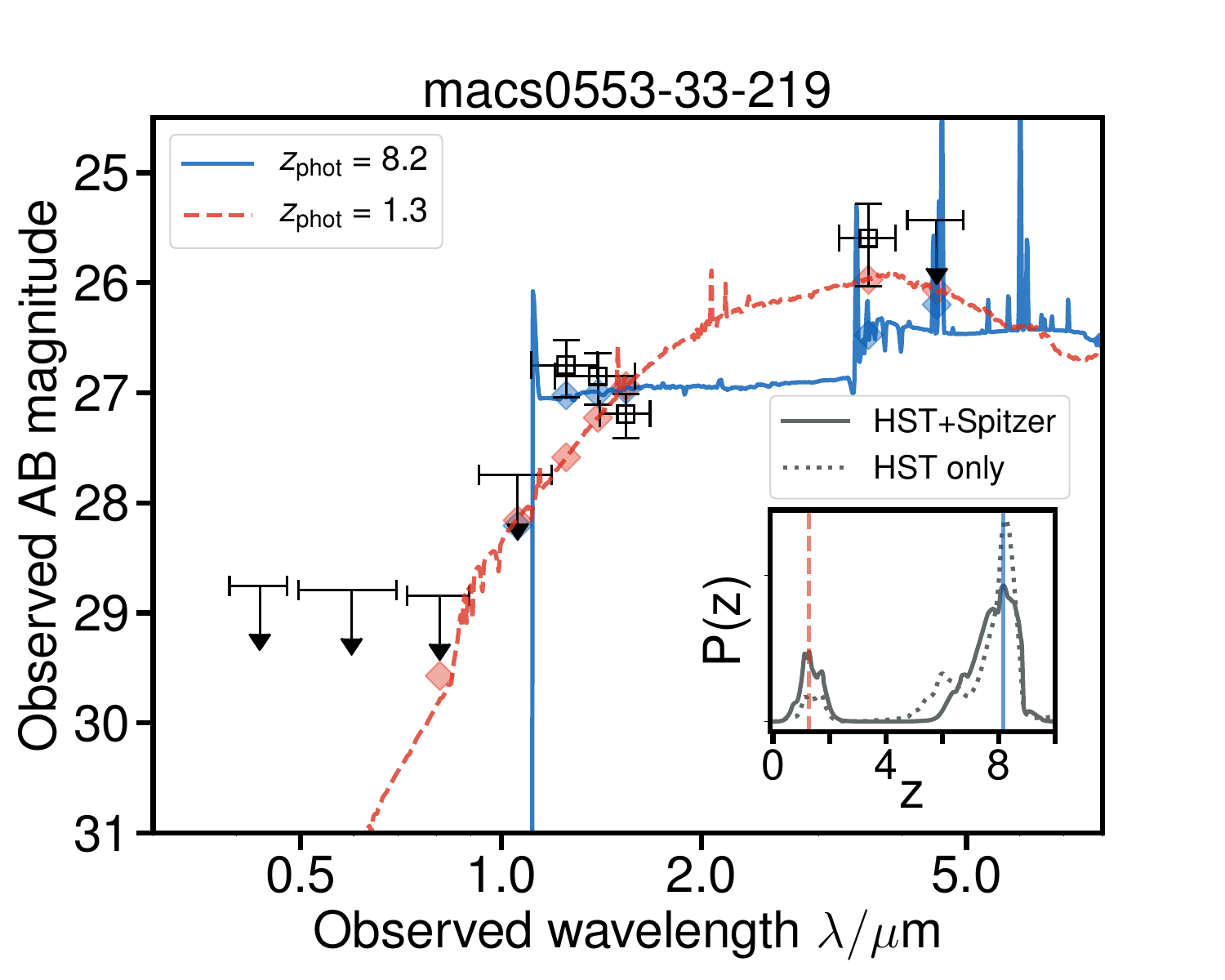}
    \end{subfigure}
   \begin{subfigure}
        \centering
        \includegraphics[width=6.4cm]{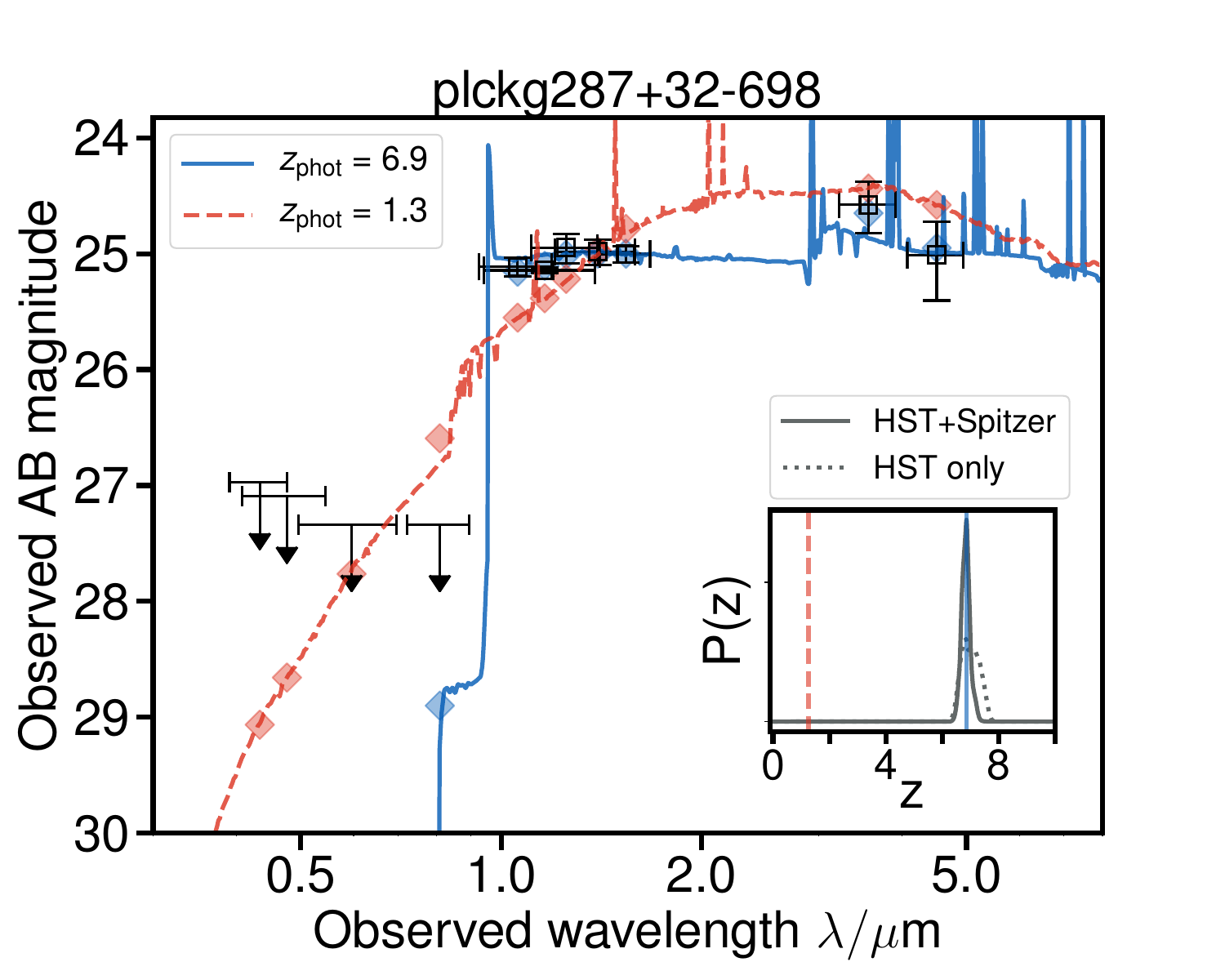}
    \end{subfigure}
    \hspace{-0.7cm}
    \begin{subfigure}
        \centering
        \includegraphics[width=6.4cm]{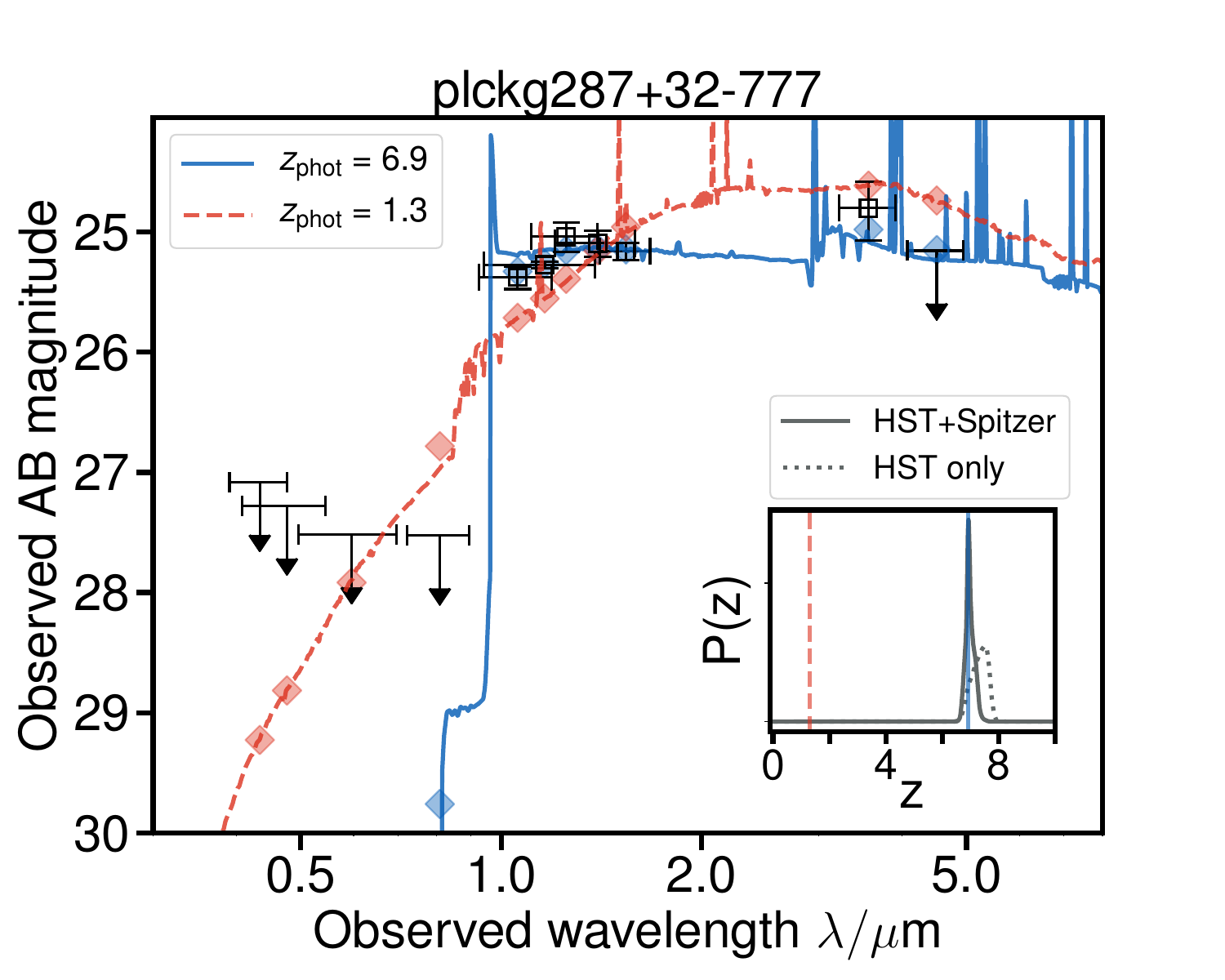}
    \end{subfigure}
    \hspace{-0.7cm}
    \begin{subfigure}
        \centering
        \includegraphics[width=6.4cm]{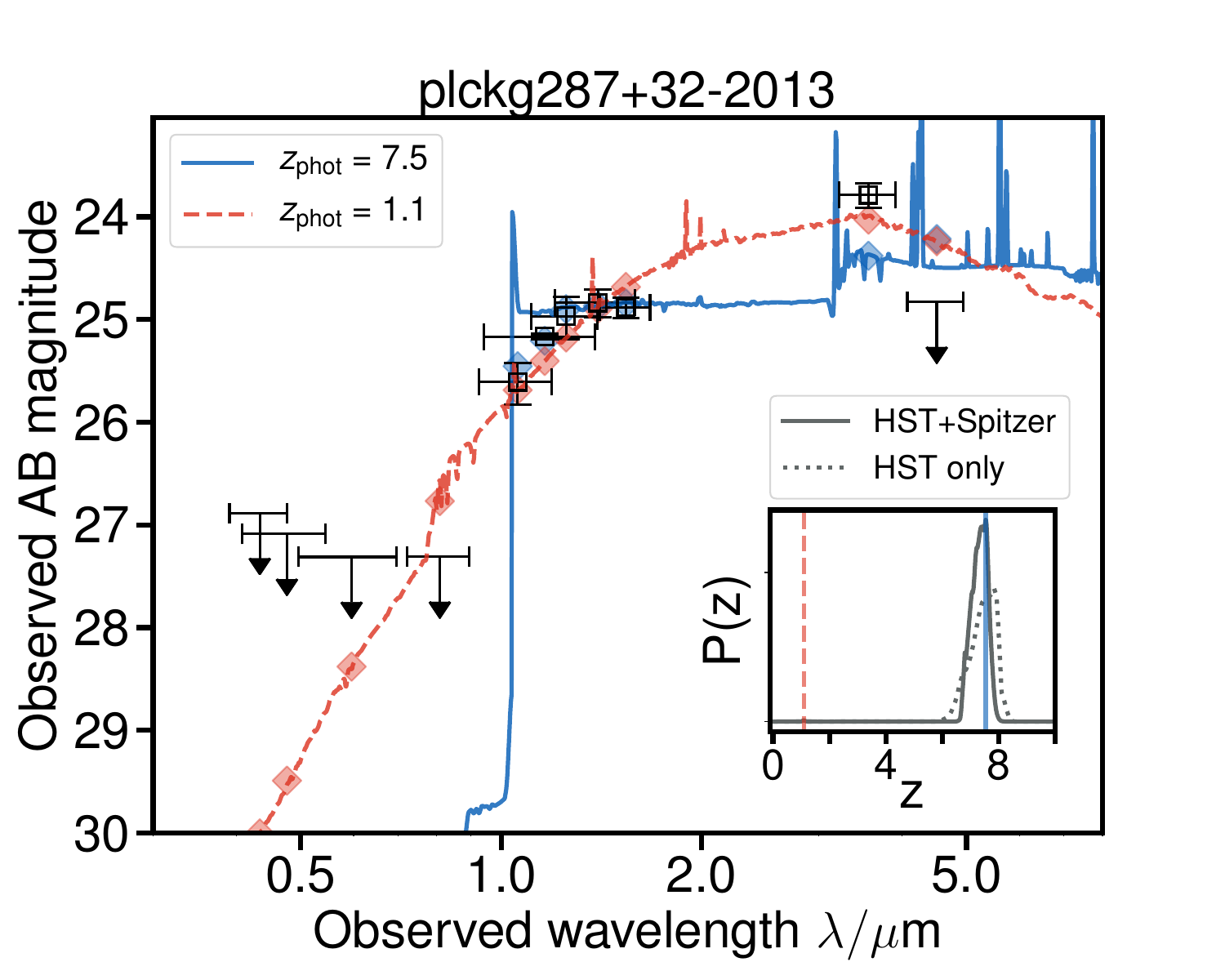}
    \end{subfigure}
    \begin{subfigure}
        \centering
        \includegraphics[width=6.4cm]{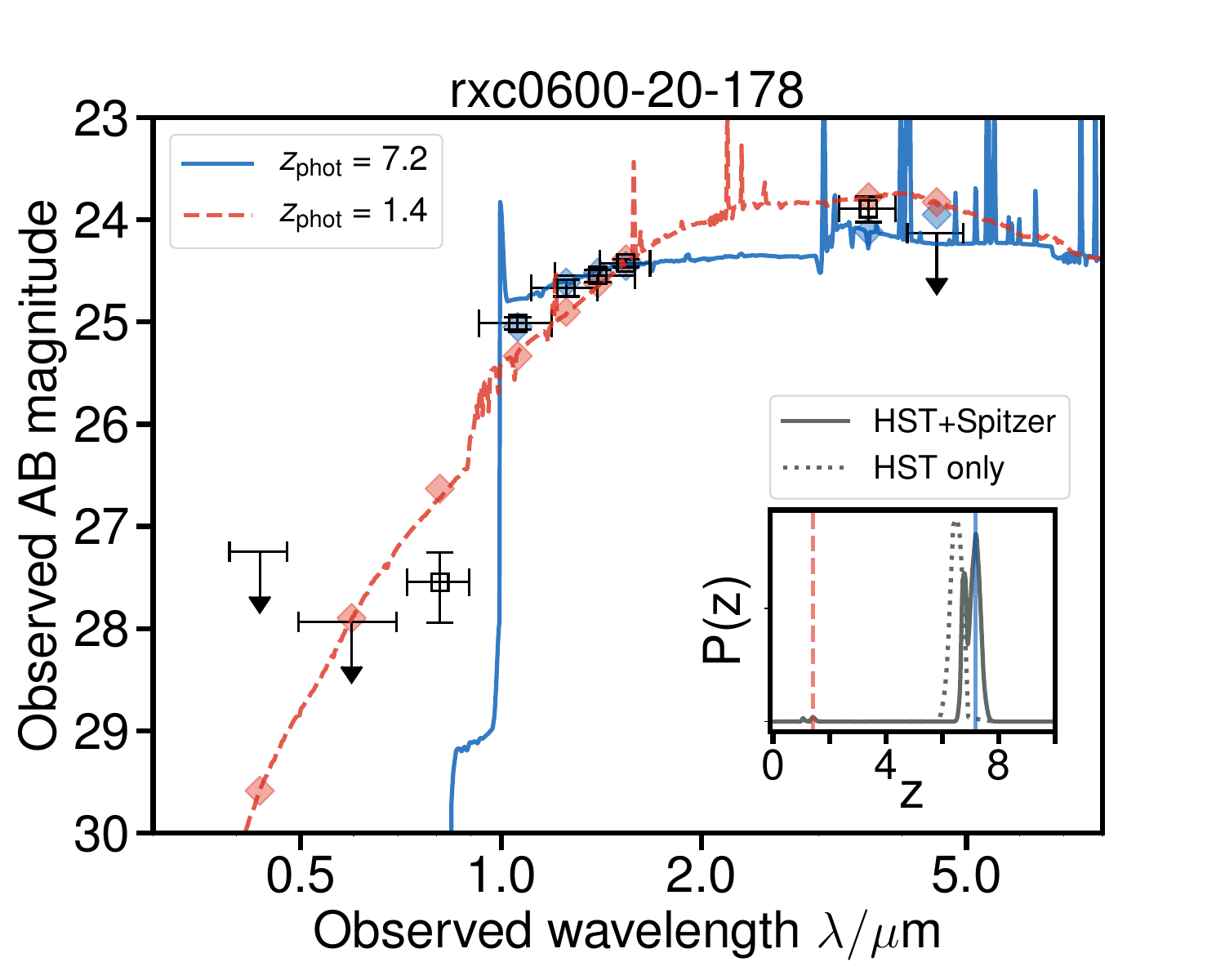}
    \end{subfigure}
    \hspace{-0.7cm}
    \begin{subfigure}
        \centering
        \includegraphics[width=6.4cm]{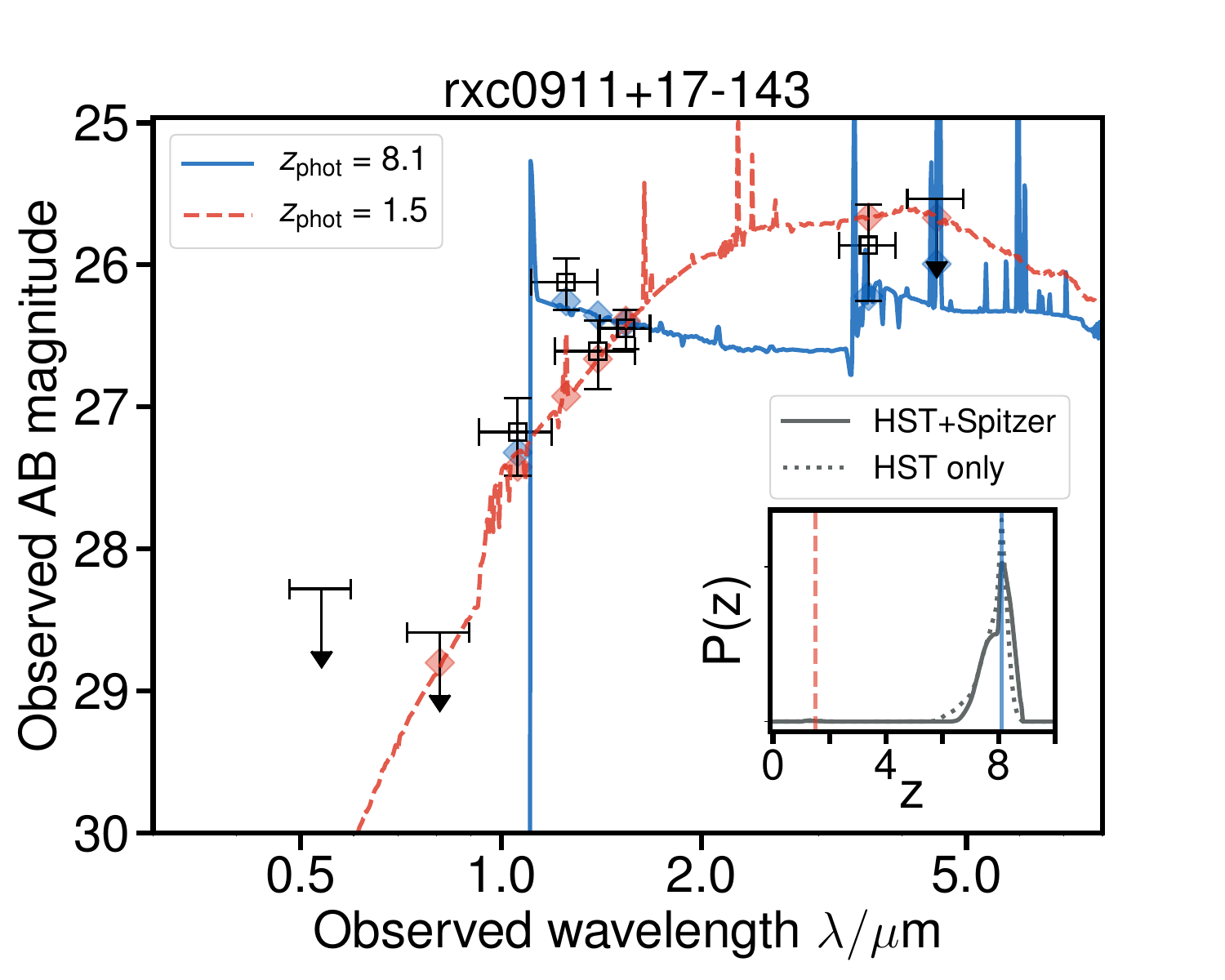}
    \end{subfigure}
    
    \caption{The SEDs and best-fit templates from the Method A SED fitting described in Section \ref{sedfit} for each of the 11 galaxies which have both $z \geq 6.5$ and at least one detection in \emph{Spitzer}/IRAC. Data are shown as black squares with error bars, and predicted photometry from models are shown as red and blue translucent diamonds. Upper limits and error bars for HST and Spitzer are at the 3-$\sigma$ level. Blue template is the object's best-fit high redshift ($z>4$) template; red template is the object's best-fit template at the secondary peak in the redshift PDF. \textbf{Inset: }Redshift PDF; dotted line is using HST only (as in \citealp{Salmon2020}) and solid line is PDF using \emph{HST} and \emph{Spitzer} fluxes. }
    \label{fig:bestfitseds}
\end{figure*}
Throughout this work, we will refer to two separate methods for estimating physical properties of galaxies. Method A is our primary routine which we use in order to compare to previous similar works (e.g., \citealp{Huang2016b}). We mainly refer to results from Method A in this work (e.g., the values in Table \ref{tbl-3}, results in Figures \ref{fig:bestfitseds} and \ref{fig:distributions}). Method B is a secondary routine which we will use for comparison purposes in Figure \ref{fig:histograms}, as well as discussion of biases and uncertainties in our analysis (Section \ref{biasandunc}), and in the exploration of the six objects we discuss in detail in Section \ref{results}. We report results from both analyses in the attached catalog (see Section \ref{catalogsect}.) 

\subsection{Method A}
In our primary SED fitting method, we first fit for redshift, and then use the resultant redshift posterior when fitting for other properties. We start by calculating colors from HST and IRAC imaging to estimate a redshift probability distribution function (PDF) for each source using the redshift estimation code Easy and Accurate Redshifts from Yale (EA$z$Y, \citealp{Brammer2008}). This code compares the observed SEDs to a set of stellar population templates. Using linear combinations of a base set of templates from \citeauthor{Bruzal2003} (\citeyear{Bruzal2003}, BC03) and no priors on magnitude, EA$z$Y performs \mbox{$\chi^2$} minimization on a user-defined redshift grid, which we define to range from \mbox{$z=0.1-12$} in linear steps of \mbox{$\delta z=0.1$}, and computes a PDF from the minimized \mbox{$\chi^2$} values. We use this PDF to calculate best-fit redshift and uncertainties, but opt for a slightly different process to calculate stellar properties.

To calculate stellar properties, we sample from the redshift PDF (we do not fix a redshift) and our photometry assuming a normal error distribution 1000 times, each time finding a best fit template, again with EA$z$Y. In order to be able to extract physical properties associated with each template, we do not allow linear combination of templates during this stage of SED fitting. We use a set of $\sim$2000 stellar population synthesis templates from the updated BC03 templates from 2016\footnote{http://www.bruzual.org/~gbruzual/bc03/}, assuming the following: a Chabrier initial mass function (IMF, \citealp{Chabrier2003}) between 0.1 and 100 $M_{\odot}$, a metallicity of 0.2 $Z_{\odot}$, a constant star formation history (SFH), a Small Magellenic Cloud (SMC) dust law \citep{Prevot1984} with $\rm{E_{stellar}}$(B-V)=$\rm{E_{nebular}}$(B-V) with step sizes of $\Delta$E(B-V)=0.05 for 0-0.5 mag and 0.1 for 0.5-1 mag. We allow formation age (the age from when the first stars in the galaxy formed) to range from 10 Myr to the age of the universe at the redshift of the source. In Section \ref{biasandunc}, we discuss the motivations behind these assumptions and the biases and uncertainties that come from our choices. 

We also include nebular emission lines and continuum to account for the role these can have in contributing to broadband fluxes \citep{Schaerer2010,Smit14}. To do this, we first calculate hydrogen recombination line strength following the relation from \cite{Leitherer1995}, scaling from integrated Lyman-continuum flux, and then follow the strengths determined with nebular line ratios by \cite{Anders2003}. We also include Lyman-$\alpha$, calculating expected strengths using the ratio with H-$\alpha$ and assuming a Case B recombination in \cite{Brocklehurst1971} with a Ly-$\alpha$ escape of 20\%. 

\subsection{Method B}
For our secondary method of estimating galaxy properties, we use the program Bayesian Analysis of Galaxies for Physical Inference and Parameter EStimation (BAGPIPES, \citealp{Carnall2018}).
Instead of the default BC03 models, we implemented the Binary Population and Spectral Synthesis (BPASS v2.2.1, \citealp{Eldridge2009}) templates,
reprocessed to include nebular continuum and emission lines using the photoionization code \texttt{CLOUDY} c17.01 \citep{Ferland2017}. 
BAGPIPES fits redshift in parallel with physical properties of galaxies using the MultiNest nested sampling algorithm \citep{Feroz2008,Feroz2009,Feroz2013} to navigate the multi-dimensional space of physical parameters. 
For this work, we implement an exponential SFR with flexibility to rise or decline at any rate, or stay constant: 
\begin{align}\label{eqn:SFH}
    SFR(t) \propto e^{-t / \tau}, \\
    \tau = \frac{t_{obs} - t_{form}} {\tan(R \pi / 2)} ,
\end{align}
for a galaxy observed at time $t_{obs}$ that began forming stars at time $t_{form}$.
The parameter $R$ dictates the rate of increase, ranging from $-1$ (maximally old burst), through negative fractional values (declining), 0 (constant), positive (increasing), to $+1$ (maximally young burst). 
We use the BPASS IMF {\tt imf135\_300}: Salpeter slope $\alpha = -2.35$ between 0.5 and 300 $M_{\odot}$ and shallower $\alpha = -1.3$ for lower mass stars 0.1 -- 0.5 $M_{\odot}$.
Dust in BAGPIPES assumes a functional form described by the Calzetti Law \citep{Calzetti2000}, 
and we assign twice as much dust around HII regions as in the general ISM in the galaxy's first 10 Myr. 
We allow dust extinction to range from $A_V = 0$ to 3 magnitudes. 
We vary metallicity in log space from 0.005 to 5 $Z_{\odot}$ and allow ages of formation from 1 Myr to the age of the universe. 
These parameters are summarized in Table \ref{tbl-4}, and biases and uncertainties are discussed in Section \ref{biasandunc}.

We reprocess the BPASS stellar continuum spectra using \texttt{CLOUDY}. Hydrogen and helium scale factors, as well as other model details, are listed in \cite{Eldridge2009}. We allow the ionizaton parameter log($U$) to vary from $-4$ to $-1$, allowing a very large dynamic range of H-$\beta$[OIII]$\lambda$4959,5007\AA\ (EW can be $>10000$\AA\ for the most extreme ionization conditions).

\begin{figure*}
        \centering
        \includegraphics[width=14.5cm]{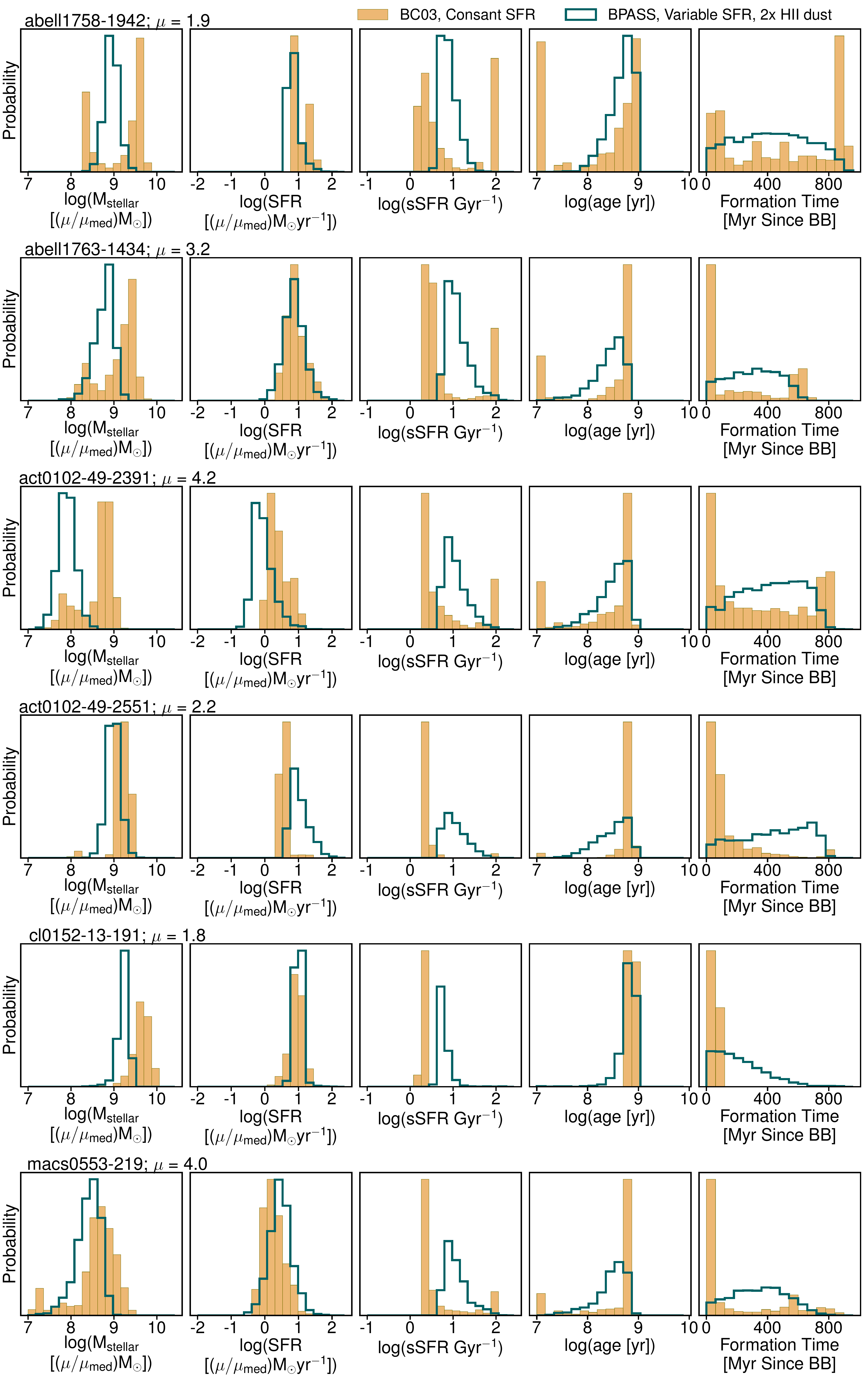}
     
\end{figure*}

\begin{figure*}[ht!!]
        \centering
        \includegraphics[width=14.5cm]{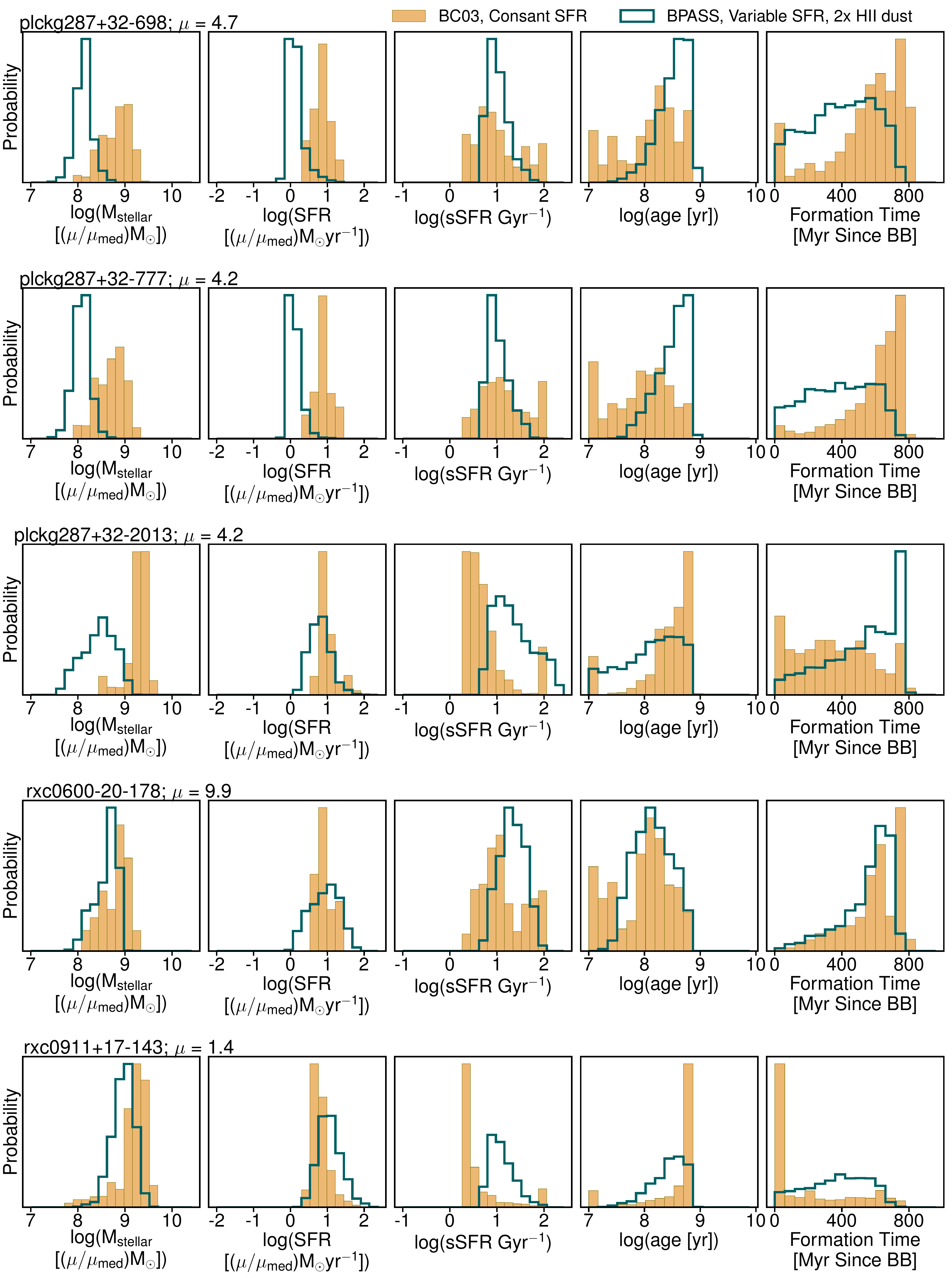}
     \caption{\label{fig:SEDfitting}Results of SED fitting for the 11 objects in the sample which had both a redshift of at least $z \geq 6.5$ and at least one detection in \emph{Spitzer}/IRAC. In yellow/tan filled histograms, the distribution of stellar mass, SFR, sSFR, formation age, and formation time resulting from Method A, and dark green open histograms show the distribution of the same properties resulting from Method B. These are described in full in Section \ref{sedfit}.}
    \label{fig:histograms}
     
\end{figure*}


\subsection{Biases and Uncertainties}\label{biasandunc}
The biases and uncertainties associated with SED fitting have been well-documented (e.g., \citealp{Papovich2001,Shapley2001,Lee2009,Salmon2015,Carnall2018,Leja2019}), and we further explore in this section considerations relevant to properties of our sample presented in this work.
\subsubsection{Star Formation Histories}
While there have been studies on the effect of SFH on inferred galaxy properties at lower redshift ($z<2$, e.g., \citealp{pacifici2016,Carnall2018,Leja2019,Sidney2020}), few studies have focused on the very high redshift regime we consider. The lower-redshift studies show that a bias can be introduced in stellar mass, SFR and age from a choice of any parameterized SFH model (e.g., exponential or constant), due to the intrinsically bursty and stochastic nature of star formation. Specifically, \cite{Carnall2019} find that stellar mass, SFR and mass-weighted age vary with the choice of SFH model for mock photometry at $z=0$ by at least 0.1, 0.3 and 0.2 dex, respectively. Furthermore, they find that, generally, photometric data alone cannot discriminate between SFH models. 
While the need to constrain SFHs may be somewhat alleviated at very high redshifts, as the universe has only had $\sim$750 Myr to form stars at $z\sim8$, the very beginning of SFHs leave little imprint on a galaxy's SED, making it difficult to fully constrain. For this reason, we distinguish between 
formation age and mass-weighted age.
We define formation age as $t_{obs} - t_{form}$,
or the age of a galaxy observed at time $t_{obs}$ since it first started forming stars at time $t_{form}$.
While this is a useful parameter for constraining the beginning of galaxy formation and reionization, we also report mass-weighted age, as this is a more robustly constrained parameter \citep{Leja2019}. 
Mass-weighted age is defined as:

\begin{align}\label{eqn:mwage}
    {\rm age}_{MW} &= t_{obs} - t_{MW} \\
    t_{MW} &=  \frac{\int_{t_{form}}^{t_{obs}} t \; SFR(t)\; dt}{\int_{t_{form}}^{t_{obs}} SFR(t)\; dt}.
\end{align}

\subsubsection{Metallicity and Dust}
In our Method A SED fitting template library, we assume a metallicity of 0.2$\rm{Z_\odot}$. While there is evidence that this assumption is appropriate for at least some of our sample \citep{Jones2020}, we still test the effects of these biases by fitting our data to templates using several dust laws and metallicities, and find that changing the dust law choice from SMC to Milky Way biases stellar masses high by $\sim0.5$ dex, and that even large changes in metallicity introduce subdominant biases, $<0.1$ dex. In Method B, we allow a range of metallicities and attenuations in order to remain agnostic to this issue. 
\subsubsection{Nebular Emission}
It has been shown that emission lines can contribute significantly to broadband flux, and several recent studies have shown that these emission lines can be more extreme than previously thought. For example, the combination of H-$\beta$ and [OIII] can sometimes be up to $\sim3000$\AA\ in their rest-frame equivalent width \citep{Labbe2013,Finkelstein2013,Smit2015,Roberts-Borsani2020}, and are regularly indirectly observed to be $\sim$700\AA\ \citep{Endsley2020}. We have included nebular emission in our templates, as described in Section \ref{sedfit}. With our observational setup, the most relevant contaminating lines are [OIII] and H-$\beta$, as these are the likely strongest emission lines in our observed-frame window. These lines are observed in our templates from Method A to span a dynamic range from 0 to $\sim2000$\AA\ in rest-frame equivalent width. This range allows for extreme line emitting galaxies to be identified for all but the rarest objects. However, there is a degeneracy in IRAC colors between extreme line-emitting galaxies at $z\sim8$ and those with a steep Balmer/$\rm{D_n}$(4000\AA) break (hereafter, Balmer break), signifying an evolved stellar population. Since we allow for both in our templates in Method A, and allow ionization parameter log(U) to vary to extreme values in Method B, where [OIII] can have EWs upwards of 10000\AA, our approach makes us agnostic to this issue. The degeneracy between a Balmer break and strong [OIII]+H-$\beta$ emitters, however, cannot be truly disentangled without secure measurements of [OIII]+H$\beta$ EW with \textit{JWST}, and in some cases ($z>9.1$), indirectly with spectroscopic redshifts, to determine in which IRAC band the emission lines fall. 
\subsubsection{Binary Stars}
Lastly, it has been shown that the inclusion of binary stars in the creation of stellar population synthesis templates can influence the results \citep{Eldridge2008}. For this reason, we use BPASS templates in Method B. We find that the change from BC03 to BPASS templates alone does not significantly change our results in terms of the average redshifts and stellar properties of our candidates, though considerable changes can be induced for individual candidates. 


\subsubsection{Differences in SED Fitting Methodology}
Due to the fact that we are varying assumptions in modeling as discussed above as well as using two different methods of fitting models to our data, the differences must be carefully interpreted. In Method A, we are creating a grid of best-fit solutions after each iteration of sampling from photometry. The results of this method rely heavily on the template set that we have generated -- specifically, how age is sampled, and what ionization or dust conditions are allowed. Taking only the best-fit on each iteration may not take into account solutions that are nearly as good as the best fit solution (e.g., the ``second-best fit''). In turn, this may affect our results by not including all ``good" fits to the data. We believe this may be part of the reason that our results from Method A are, in general, less broad and sometimes bimodal, rather than a broader, unimodal distribution like those seen in Method B. 

Method B uses Nested Sampling, which, similar to Markov Chain Monte Carlo techniques, is a robust way to calculate posteriors for distributions that may be multimodal or may have pronounced degeneracies. Rather than sampling from photometry, BAGPIPES explores an $n$-dimensional parameter space, each time judging the goodness of fit and, taking into account the input priors, outputs a marginalized posterior for all $n$ parameters that are being fit. This means that generally, a smoother distribution of each parameter is found, and nearly all ``good" solutions for the data are reported in the posteriors. 

While these differences in methodology affect our results for individual galaxies, we note that the overall sample distributions (stellar mass, SFR, sSFR) do not change dramatically, with the exception of formation age (see Section \ref{results}). This is likely because formation age is a difficult parameter to constrain, and due to our assumption of a constant SFH in Method A. 

\section{Lens Modeling}\label{magmaps}
\subsection{Magnification Maps}
Many of our derived properties are unaffected by lensing,
including age, dust extinction, metallicity, sSFR, and emission line strengths.
Other properties (stellar mass, SFR, $\rm{M_{UV}}$) 
must be corrected for lensing magnifications.
To do so,
we use magnification maps provided by RELICS lens modeling teams\footnote{archive.stsci.edu/prepds/relics/}. The maps we utilize in this work are from one of three types of lens models: $\texttt{Lenstool}$ \citep{jullo2009}, Glafic \citep{Oguri2010}, and light-traces-mass (LTM, e.g., \citealp{zitrin13}). Several papers have been published describing the lens models \citep{Cerny2018,Acebron2018,Cibirka2018,PaternoMahler2018,Acebron2019a,Mahler2019,Acebron2019b,Okabe2020}. 
All models used in this work are available on MAST for public use.
Not every model has been detailed in a publication, but the general methods are very similar to those described in the papers listed here. 

For each method, a routine bootstrapping of lensing constraints is performed to create 50--100 models,
each yielding a magnification estimate given a lensed galaxy's position and redshift.
We adopt the Method B median redshifts, extract the 50--100 magnification estimates, then take the median as the estimate from each method.
For clusters modeled by 2 or 3 methods, we take the average of those median estimates as our final magnification estimate for that galaxy.

The lens models provide magnification estimates for 150 of our 207 high-z candidates.
The median of these 150 estimates is $\mu = 2.9$, and 127 (85\%) of those have $\mu < 10$.
Only 6 have $\mu > 30$, the highest being $\mu = 96$ for plckg287+32-2457, an average of 107 and 85 from GLAFIC and LTM, respectively.
We have reason to doubt individual magnification estimates of $\mu > 30$ \citep{Menenghetti2017}, 
but the average of medians lends somewhat more confidence to the few higher estimates reported here.

For some RELICS clusters, no strong lensing models are available, as no multiple images were identified in the \emph{HST} imaging.
In those cases, we adopt a nominal magnification of $\mu = 3$ for all high-z candidates.
This is roughly the median of the 150 estimates and within a factor of 3 for 125 (83\%) of those.

Given magnification estimates for all 207 high-z candidates,
we divide our derived mass, SFR, and luminosity
by these magnifications before reporting the results in Table \ref{tbl-3}.
We also report the 68\% confidence limits of each magnification estimate,
but we do not add these to the reported uncertainties of the derived properties.
This is so that the reader can easily use their own magnification and uncertainty measurements.


The models have varying numbers of multiple-image lensing constraints, making them not all equally reliable. 
For some clusters, no multiple images were identified for strong lensing analysis.
In those cases, we relied on ``blind'' LTM models (in contrast with the regular LTM models mentioned above where constraints are available) with a mass-to-light normalization typical of other clusters\citep{zitrin12}.
For these blind LTM models, the (larger) uncertainties follow from the uncertainties in typical mass-to-light normalizations.


\subsection{Source Plane Modeling}
 We perform source-plane modeling for the 11 galaxies highlighted in this manuscript, and show the results in Figure \ref{fig:resids}. Closely following the methods of \cite{Yang2020}, we use the publicly available code \texttt{Lenstruction} (based on \texttt{Lenstronomy}\footnote{https://github.com/sibirrer/lenstronomy}) to forward model the high-$z$ galaxies using their appearance in the image plane to predict their morphology and size in the source plane. \texttt{Lenstruction} takes into account distortion from lensing and the instrument PSF, and works to estimate pixel noise and remove the background flux level. We assume Glafic global magnification models when modeling each of the galaxies. We modeled each source as a singly-imaged galaxy, and we will explore this work further in Neufeld et al., in prep. 

\section{Results}\label{results}

\begin{figure}
    \centering
    \includegraphics[width=9cm]{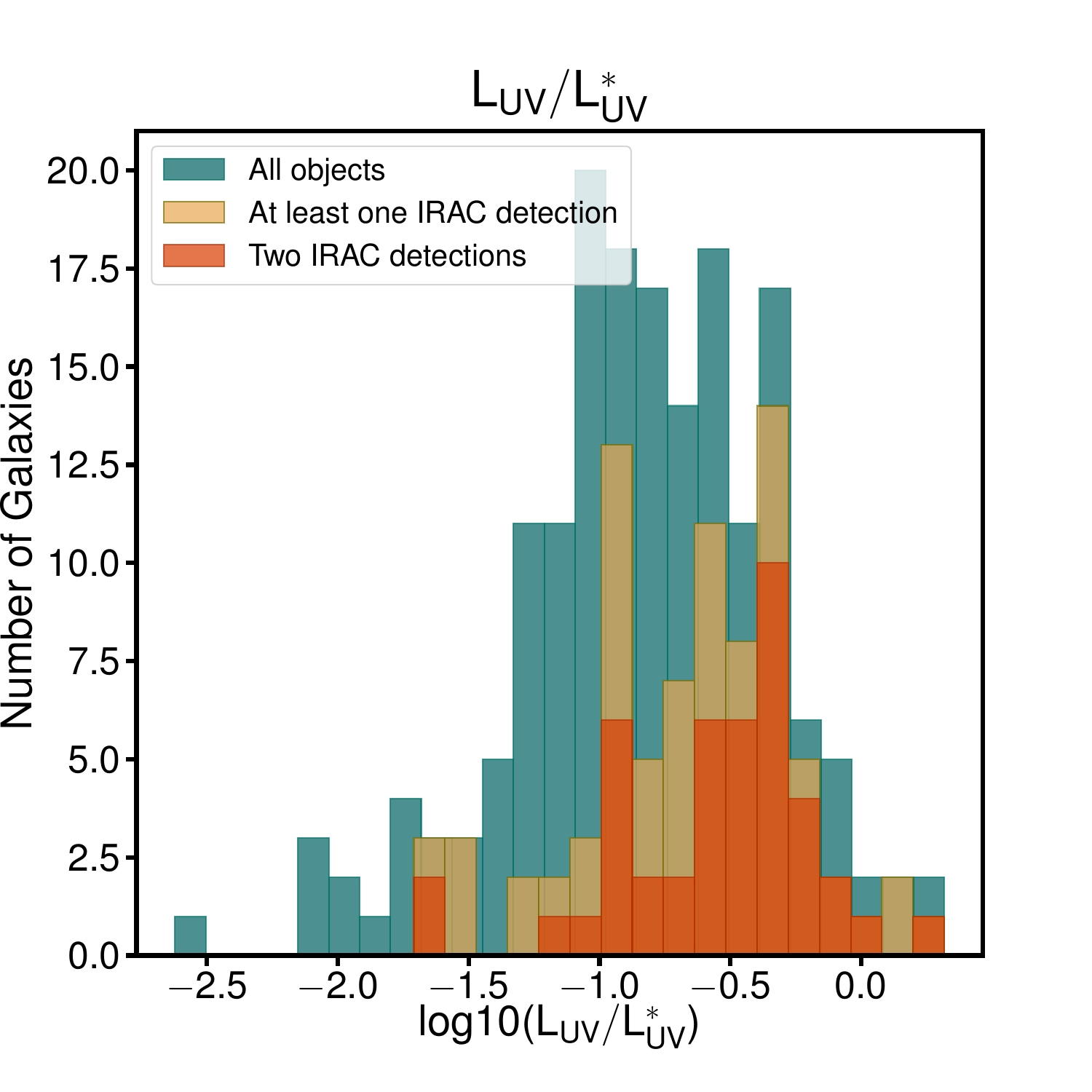}
    \caption{The distribution of $\rm{L_{UV}/L_{UV}^*}$, intrinsic luminosity normalized by the characteristic luminosity for that object's redshift, for (in teal/blue) all 207 galaxies in the sample, (in tan) galaxies that were detected at least once by \emph{Spitzer}/IRAC, and (in orange) galaxies that were detected twice in \emph{Spitzer}/IRAC. A large majority of all three subsets of the sample are $L_{UV}/L^*_{UV}<1$, with a few objects at $L_{UV}/L_{UV}^*\sim2$.}
    \label{fig:lstar}
\end{figure}

\begin{figure*}
    \begin{subfigure}
        \centering
        \includegraphics[width=6cm]{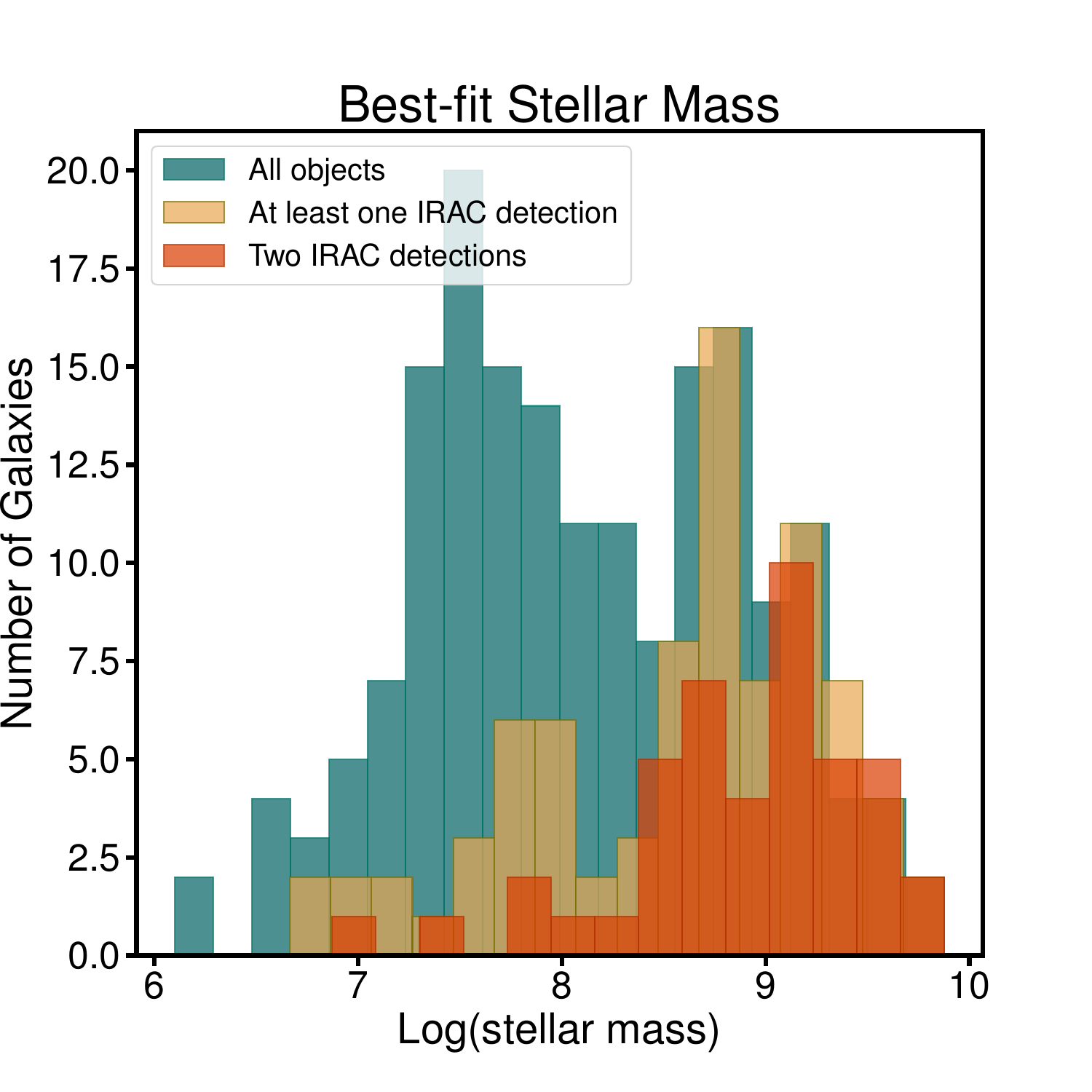}
    \end{subfigure}
    \begin{subfigure}
        \centering
        \includegraphics[width=6cm]{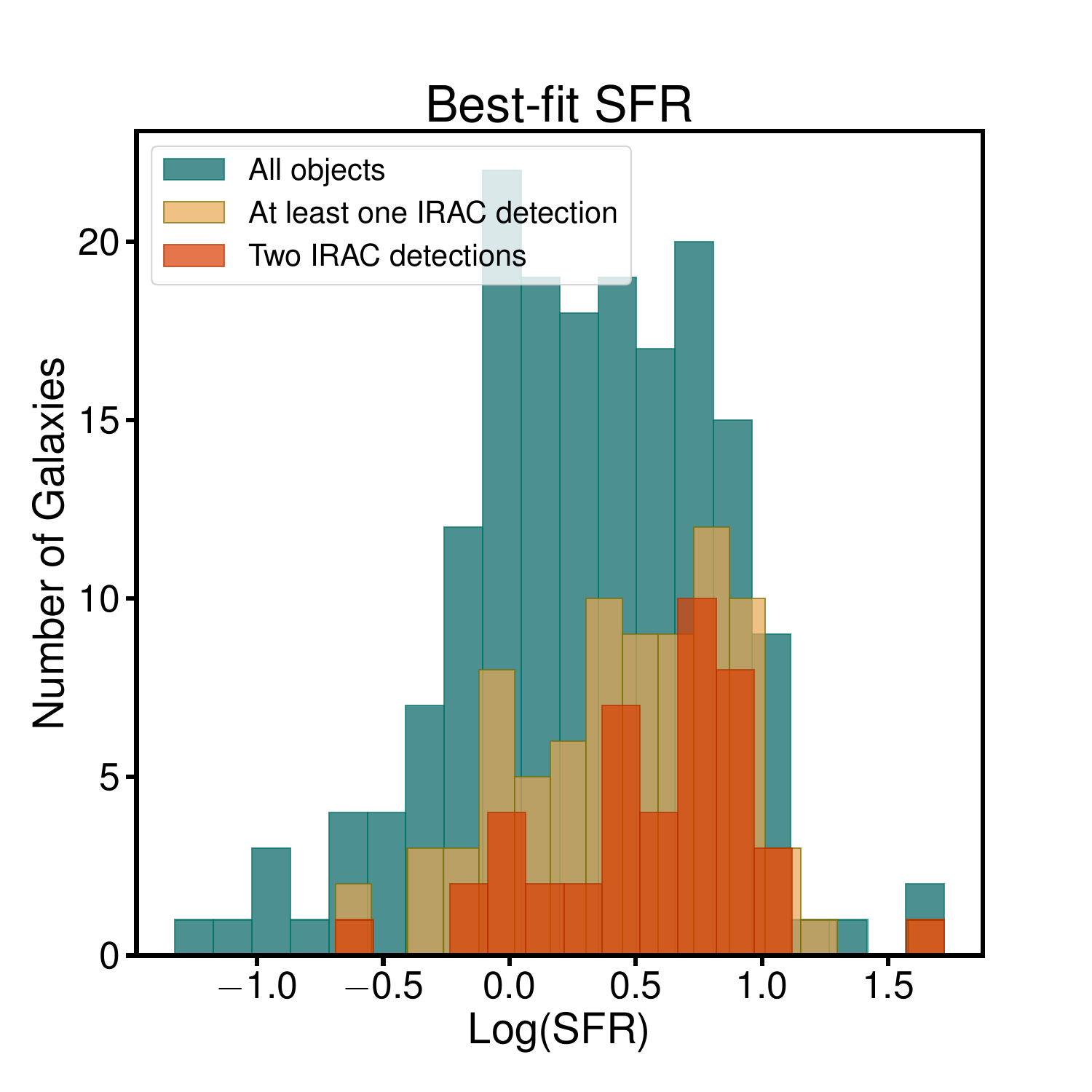}
    \end{subfigure}
    \begin{subfigure}
        \centering
        \includegraphics[width=6cm]{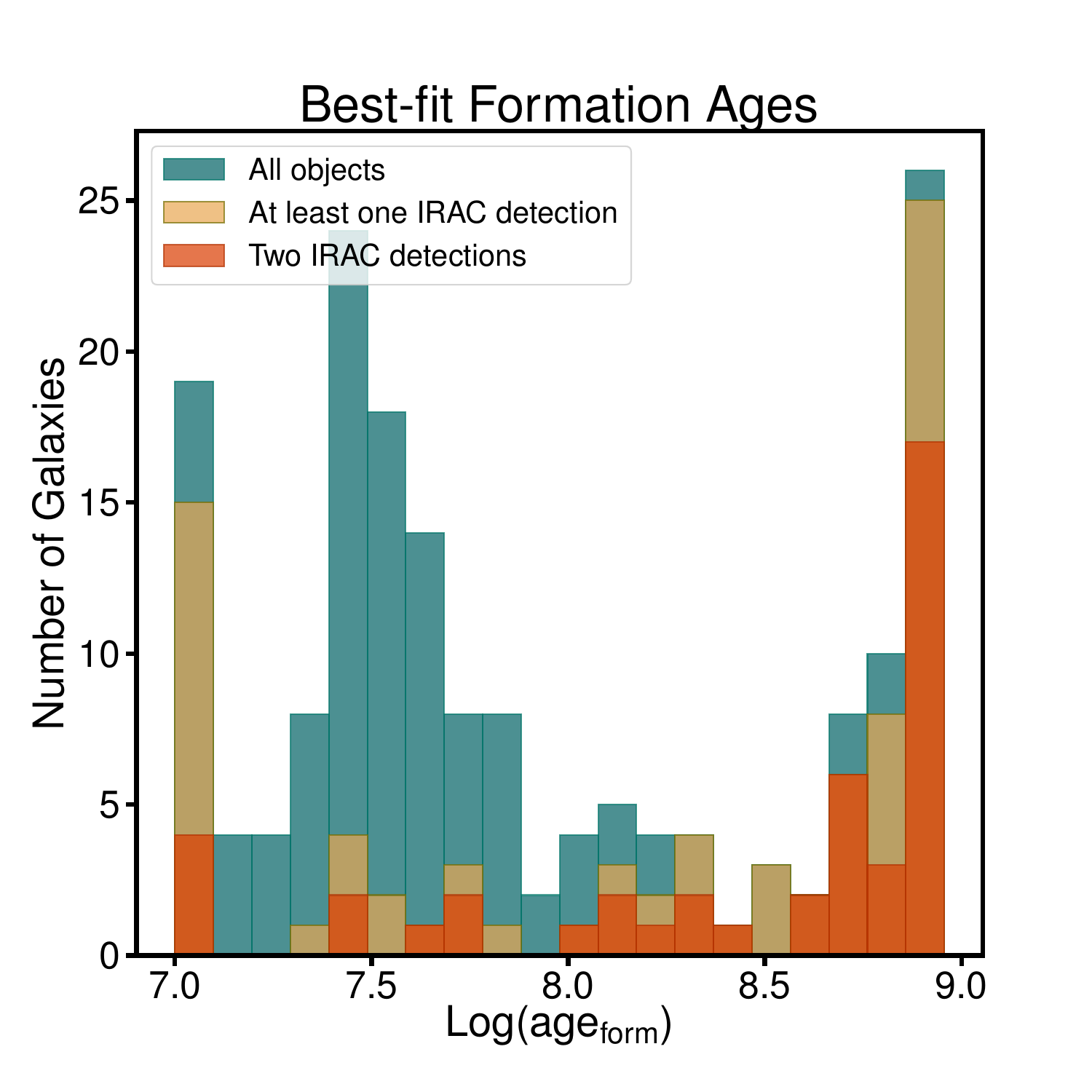}
    \end{subfigure}
    \label{fig:distributions}
    \caption{Histograms of best-fit physical parameters derived by Method A.
    In all panels, teal/blue is all 207 galaxies in the sample,
    tan is galaxies with at least one \emph{Spitzer}/IRAC detection, 
    and orange is galaxies with two \emph{Spitzer}/IRAC detections. \textbf{Left: }Best-fit log stellar mass distribution ($M_\odot$). Detected galaxies fall on the higher stellar mass end, showing a bias of \emph{Spitzer} being able to detect higher stellar mass objects due to brighter rest-frame optical fluxes. 
    \textbf{Center: }Best-fit log(SFR) distribution ($M_\odot$/yr). Similar to stellar mass, higher SFR galaxies are more often detected than lower SFR (due to a correlation between M* and SFR). 
    \textbf{Right: }Best-fit log(age) distribution, where age here is age of formation in years: $t_{obs} - t_{form}$.
    Galaxies detected in \emph{Spitzer} tend to be older. Since we are assuming a constant SFH, this makes sense because older galaxies will have had more time to form more mass, an effect likely due to both the intrinsic properties of the galaxies we detect and our assumption of a constant SFH, which requires more massive galaxies to be older.}
\end{figure*}


\begin{deluxetable*}{lccccccccccccccc}
\tabletypesize{\footnotesize}
\tablecaption{\label{tbl-2} Spitzer Photometry of Selected Galaxies} 
\tablewidth{0pt}
\tablehead{
\colhead{Object ID} &  \colhead{R.A.} & \colhead{Dec.} & \colhead{$F160W^{\tablenotemark{1}}$}& \colhead{$[3.6]^{\tablenotemark{2}}$} & \colhead{$R_{3.6}^{\tablenotemark{3}}$} & \colhead{$[4.5]^{\tablenotemark{2}}$} & \colhead{$R_{4.5}^{\tablenotemark{3}}$}\\  \colhead{}  & \colhead{(deg.)} & \colhead{(deg.)} & \colhead{(mag)} & \colhead{(mag)} &  \colhead{} & \colhead{(mag)} & \colhead{} 
}
\startdata
abell1758-1942 & 203.2001098 & 50.5185167 & $24.82 \pm 0.05$ & $24.03 \pm 0.17$ & 0.1 & $> 25.08$ & 0.1\\
abell1763-1434 & 203.8333744 & 40.99017930 & $26.17 \pm 0.08$ & $25.74 \pm 0.49$ & 0.3 & $24.75 \pm 0.26$ & 0.3\\
act0102-49-2391 & 15.72313210 & -49.2393723 & $26.26 \pm 0.14$ & $25.29 \pm 0.21$& 0.2 & $26.03 \pm 0.39$ & 0.3 \\
act0102-49-2551 & 15.72568030 & -49.232616 & $26.25 \pm 0.12$ & $25.03 \pm 0.16$ & 0.1 & $25.49 \pm 0.20$5 & 0.1\\
cl0152-13-191 & 28.17164110 & -13.9734429 & $27.31 \pm 0.16$ & $24.69 \pm 0.15$ & 0.4 & $25.04 \pm 0.25$ & 0.5\\
macs0553-33-219 & 88.3540349 & -33.6979484 & $27.19 \pm 0.18$ & $25.60 \pm 0.31$ & 0.5 & $26.41 \pm 0.65$ & 0.5\\
plckg287+32-2013  & 177.6877971 & -28.0760864 & $24.88\pm0.09$ & $23.79 \pm0.11$ & 0.7 & $25.24\pm0.43$ & 0.7\\
plckg287+32-698 & 177.7049678 & -28.0707102 & $25.00\pm0.07$ & $24.58\pm0.20$ & 0.5 & $25.01\pm0.29$ & 0.5\\
plckg287+32-777 & 177.7199024 &-28.0715239 & $25.16 \pm 0.07$ & $24.80 \pm 0.22$ & 0.5 & $25.48 \pm 0.40$ & 0.5\\
rxc0600-20-178 & 90.0271054 & -20.1202486 & $24.42 \pm 0.04$ & $23.89 \pm 0.12$ & 0.2 &  $> 25.32$ & 0.1\\
rxc0911+17-143 & 137.7939712 & 17.7897516 & $26.45 \pm 0.13$ & $25.86 \pm 0.29$ & 0.1 & $25.72 \pm 0.36$ & 0.1 \\

\tablenotetext{1}{Observed (lensed) isophotal magnitude (\texttt{MAG}\_{\texttt{ISO}})}
\tablenotetext{2}{\textit{Spitzer/IRAC} Channels 1 and 2 magnitudes measured using T-PHOT with the same aperture as HST magnitudes and 1-$\sigma$ error. If detection is $<$ 1-$\sigma$, the 1-$\sigma$ lower limit is reported.}
\tablenotetext{3}{Covariance index for \textit{Spitzer}/IRAC channels (Section \ref{fluxextract})}
\end{deluxetable*}
\begin{deluxetable*}{lcccccccccclcccccccccc}
\tabletypesize{\footnotesize}
\tablecaption{\label{tbl-3} Photometric Redshift and Stellar Population Modeling Results from Method A of galaxies detected in [3.6] and/or [4.5] and having a best-fit redshift of $z\geq6.5$. Full catalog available at \href{http://victoriastrait.github.io/relics}{victoriastrait.github.io/relics}} 
\tablewidth{0pt}
\tablehead{
\colhead{Object ID}  &\colhead{$z_{\rm{med}}^{\tablenotemark{1}}$} &
\colhead{$\mu_{\rm{med}}^{\tablenotemark{2}}$} &
\colhead{$\rm{M}_{\rm{stellar}}^{\tablenotemark{3}}$} &   \colhead{$\rm{SFR} ^{\tablenotemark{3}}$} &   \colhead{$t_{\rm{form}}^{\tablenotemark{4}}$} & \colhead{$z_{\rm{form}}^\tablenotemark{5}$} & \colhead{$\rm{sSFR}^{\tablenotemark{6}}$} & \colhead{$\rm{E(B-V)}^{\tablenotemark{7}}$} &\colhead{$\rm{M}_{\rm{1600}}^{\tablenotemark{8}}$} &  \colhead{$\frac{L_{UV}}{L^*_{UV}}^{\tablenotemark{9}}$} \\ \colhead{} & \colhead{} & \colhead{} & \colhead{($10^9M_{\odot}$)} &   \colhead{($M_{\odot} \rm{yr}^{-1}$)} &   \colhead{(Myr)} & \colhead{} & \colhead{($\rm{Gyr}^{-1}$)} & \colhead{(mag)}  & \colhead{(mag)} &  \colhead{}
} 
\startdata

abell1758-1942 & $6.1^{+0.1}_{-0.1}$ &  $1.9^{+0.4}_{-0.2}$ & $2.0^{+1.5}_{-1.8}$ & $6.8^{+8.6}_{-0.7}$ & $404^{+500}_{-394}$ & $3.6^{+101.4}_{-1.9}$ & $0.00^{+0.00}_{-0.00}$ & $-22.1^{+0.8}_{-0.7}$ & $6.5^{+1.7}_{-0.2}$ & 2.9\\

abell1763-1434 & $8.4^{+0.4}_{-0.4}$ &  $3.3^{+2.4}_{-0.7}$ & $1.2^{+0.9}_{-1.0}$ & $5.4^{+6.9}_{-2.3}$ & $508^{+62}_{-499}$ & $2.9^{+102.1}_{-0.3}$ & $0.10^{+0.05}_{-0.05}$ & $-21.1^{+0.8}_{-0.7}$ & $47.3^{+49.4}_{-38.8}$ & 1.2\\

act0102-49-2391 & $6.9^{+0.2}_{-0.2}$ &  $4.1^{+0.7}_{-0.7}$ & $0.4^{+0.2}_{-0.3}$ & $81.5^{+3.2}_{-0.6}$ & $509^{+210}_{-497}$ & $2.9^{+85.6}_{-0.8}$ & $0.05^{+0.05}_{-0.05}$ & $-20.6^{+0.7}_{-0.8}$ & $15.0^{+39.4}_{-8.1}$ & 0.7\\

act0102-49-2551 & $6.8^{+0.1}_{-0.1}$ &  $2.2^{+0.3}_{-0.3}$ & $3.7^{+1.1}_{-1.0}$ & $9.5^{+1.2}_{-3.5}$ & $719^{+0}_{-148}$ & $2.1^{0.5}_{0.0}$ & $0.05^{+0.00}_{-0.05}$ & $-20.7^{+0.8}_{-0.7}$ & $42.1^{+9.1}_{-24.6}$ & 0.8\\

cl0152-13-191 & $6.7^{+0.2}_{-0.4}$ &  $1.8^{+0.1}_{-0.1}$ & $3.6^{+1.8}_{-1.3}$ & $7.3^{+3.8}_{-2.7}$ & $719^{+89}_{-0}$ & $2.1^{+0.0}_{-0.2}$ & $0.20^{+0.05}_{-0.05}$ & $-19.6^{+0.8}_{-0.7}$ & $46.6^{+53.1}_{-7.7}$ & 0.3\\

macs0553-33-219 & $7.6^{+0.8}_{-6.0}$ &  $2.4^{+0.3}_{-0.1}$ & $0.5^{+0.9}_{-0.4}$ & $2.1^{+3.0}_{-1.7}$ & $571^{+334}_{-480}$ & $2.6^{+11.3}_{-0.9}$ & $0.10^{+0.60}_{-0.10}$ & $-20.0^{+0.8}_{-0.7}$ & $46.9^{+56.3}_{-39.1}$ & 0.4\\

plckg287+32-698 & $6.8^{+0.2}_{-0.2}$ &  $4.3^{+0.7}_{-0.8}$ & $0.6^{+0.5}_{-0.4}$ & $5.3^{+6.0}_{-2.1}$ & $161^{+243}_{-135}$ & $8.2^{+34.7}_{-4.7}$ & $0.05^{+0.05}_{-0.05}$ & $-22.0^{+0.8}_{-0.7}$ & $8.3^{+4.0}_{-1.3}$ & 2.6\\

plckg287+32-777 & $7.0^{+0.2}_{-0.1}$ &  $4.0^{+0.7}_{-0.4}$ & $0.5^{+0.4}_{-0.3}$ & $5.5^{+5.8}_{-2.5}$ & $102^{+185}_{-84}$ & $12.5^{+50.8}_{-7.6}$ & $0.05^{+0.05}_{-0.05}$ & $-21.8^{+0.8}_{-0.7}$ & $8.0^{+2.5}_{-0.8}$ & 2.2\\

plckg287+32-2013 & $7.3^{+0.3}_{-0.3}$ &  $4.0^{+0.5}_{-0.5}$ & $1.5^{+0.6}_{-0.5}$ & $6.0^{+4.7}_{-1.0}$ & $360^{+280}_{-232}$ & $4.0^{+6.2}_{-1.6}$ & $0.05^{+0.05}_{-0.00}$ & $-22.2^{+0.8}_{-0.7}$ & $12.5^{+21.8}_{-4.2}$ & 3.2\\

rxc0600-20-178 & $7.1^{+0.2}_{-0.3}$ &  $14^{+26}_{-4}$ & $1.4^{+0.8}_{-0.8}$ & $16.1^{+15.8}_{-7.1}$ & $114^{+207}_{-95}$ & $11.3^{+46.5}_{-6.9}$ & $0.10^{+0.05}_{-0.05}$ & $-22.6^{+0.7}_{-0.8}$ & $7.8^{+1.7}_{-0.8}$ & 4.6\\

rxc0911+17-143 & $8.1^{+0.4}_{-0.6}$ &   $1.5^{+0.2}_{-0.1}$ & $1.2^{+0.8}_{-0.6}$ & $4.6^{+3.5}_{-1.7}$ & $571^{+70}_{-480}$ & $2.6^{+11.3}_{-0.3}$ & $0.05^{+0.05}_{-0.05}$ & $-20.7^{+0.7}_{-0.8}$ & $45.2^{+50.6}_{-36.1}$ & 0.8\\

\tablenotetext{1}{Median redshift and 68\% CL in PDF described in Section \ref{obsphot}}
\tablenotetext{2}{Magnification factor: we use the mean of median magnifications for each available lens model. $\mu$ is assumed in SFR,  $\rm{M}_{\rm{stellar}}$, and $\rm{M}_{\rm(1600)}$ calculations.} 
\tablenotetext{3}{Intrinsic stellar mass and SFR, assuming $\mu=\mu_{\rm{med}}$. Uncertainties include statistical 68\% CLs from photometry and redshift. To use a different magnification value, multiply the quantity by $1/f_{\mu}$, where $f_{\mu}\equiv\mu/\mu_{\rm{med}}$. }
\tablenotetext{4}{Time since the onset of star formation assuming a constant SFR}
\tablenotetext{5}{Redshift of formation, calculated from age of formation.}
\tablenotetext{6}{Specific SFR, sSFR $\equiv M_{stellar}$/SFR}
\tablenotetext{7}{Dust color excess of stellar emission. SMC dust law assumed.}
\tablenotetext{8}{Rest-frame 1600 \mbox{\normalfont\AA} magnitude assuming $\mu{\rm_{med}}$, derived from the observed F160W mag including a small template-based $k$-correction. Uncertainties include statistical 68\% CLs from photometry and redshift. To use a different magnification value, use $\rm{M}_{\rm(1600)}-2.5\rm{log}(f_{\mu}$).}
\tablenotetext{9}{Absolute UV luminosity over characteristic luminosity, assuming median redshift and $M_{1600}$.}

\end{deluxetable*}
\subsection{Sample Selection}
The goal of this work is to explore the stellar properties of $z\geq5.5$ galaxies. We start from the sample defined in \cite{Salmon2020}, which consists of 321 \emph{HST}-selected galaxy candidates with a best-fit $z\geq5.5$. Objects in this sample were required to have a median or peak in photometric redshift \mbox{$P(z)$}  at \mbox{$z \geq 5.5$} in at least one of the redshift-fitting codes used, a \mbox{$>3\sigma$} detection in F160W, stellarity of \mbox{$<98\%$}  (excluding point sources), and (Y-J) color $> 0.45$ (both to filter out lower redshift galaxies and brown dwarfs). The objects also passed an extensive visual inspection process where they were vetted for diffraction spikes, transients, detector edge noise, and other image artifacts (see \citealp{Salmon2020} for more details). 

We were able to successfully extract fluxes or flux upper limits in magnitude for 207 galaxies from \emph{Spitzer}/IRAC observations using the process described in Section \ref{fluxextract}. The remaining 114 galaxies were too crowded by bright neighboring sources, usually cluster members. Most objects for which we could not extract a \textit{Spitzer}/IRAC flux were within $\sim$1" of a cluster member or other bright ($<17$ AB mag) galaxy. For a flux to be considered reliable, we require that the residual (model - image) pixels be centered close to zero with the exception of artifacts from bright objects. The galaxies which were rejected generally showed a large over-subtraction or an obvious artifact in the residual. 

Of the objects rejected from the sample, their apparent magnitudes, luminosities, and magnifications span a similar range as that of the entire sample. From this, we conclude that we are not biasing our sample by removing these galaxies. 

Of the objects with reliable photometry, 96 have at least one IRAC detection of S/N $>1$. The overall redshift distribution did not change significantly from the \textit{HST}-only distribution (Figure 6 in \citealp{Salmon2020}), though 23 galaxies now have significant peaks in redshift at $z<4$. In our analyses, we focus mostly on our final high redshift candidate sample, but discuss the demoted objects further below. 
\subsection{Construction of Catalog}\label{catalogsect}
In the catalog\footnote{\href{http://victoriastrait.github.io/relics}{victoriastrait.github.io/relics}}, we present the photometry of the 207 objects for which we successfully extracted \emph{Spitzer} photometry, including \emph{HST} fluxes and covariance indices from T-PHOT which acts as a flag for possible blending with neighboring sources (galaxies with covariance indices $\rm{R_{[3.6],[4.5]}}>1$ should be treated with caution, see Section \ref{obsphot} for the exact definitions). Along with photometry and covariance indices, we include median magnifications and 68\% confidence limits from each lens model available on MAST, and the mean of those for any clusters with more than one model. In addition to photometry and magnification, we include the median values and upper and lower 68\% confidence limits on redshift, SFR, stellar mass, sSFR, and age from our Method A and B described in Section \ref{sedfit}. Units are described in the header of the catalog. We note that objects MACS0553-33-1014 and MACS0553-33-1016 were considered separate objects in \cite{Salmon2020}, but here we consider them the same galaxy, as their \emph{Spitzer}/IRAC fluxes are blended and they are $\sim0.6"$ apart. In the catalog, they are listed as object MACS0553-33-1014. All other objects have the same ID and coordinates as in \cite{Salmon2020}.
\begin{figure}[h!!!]
        \centering
        \includegraphics[width=9.3cm]{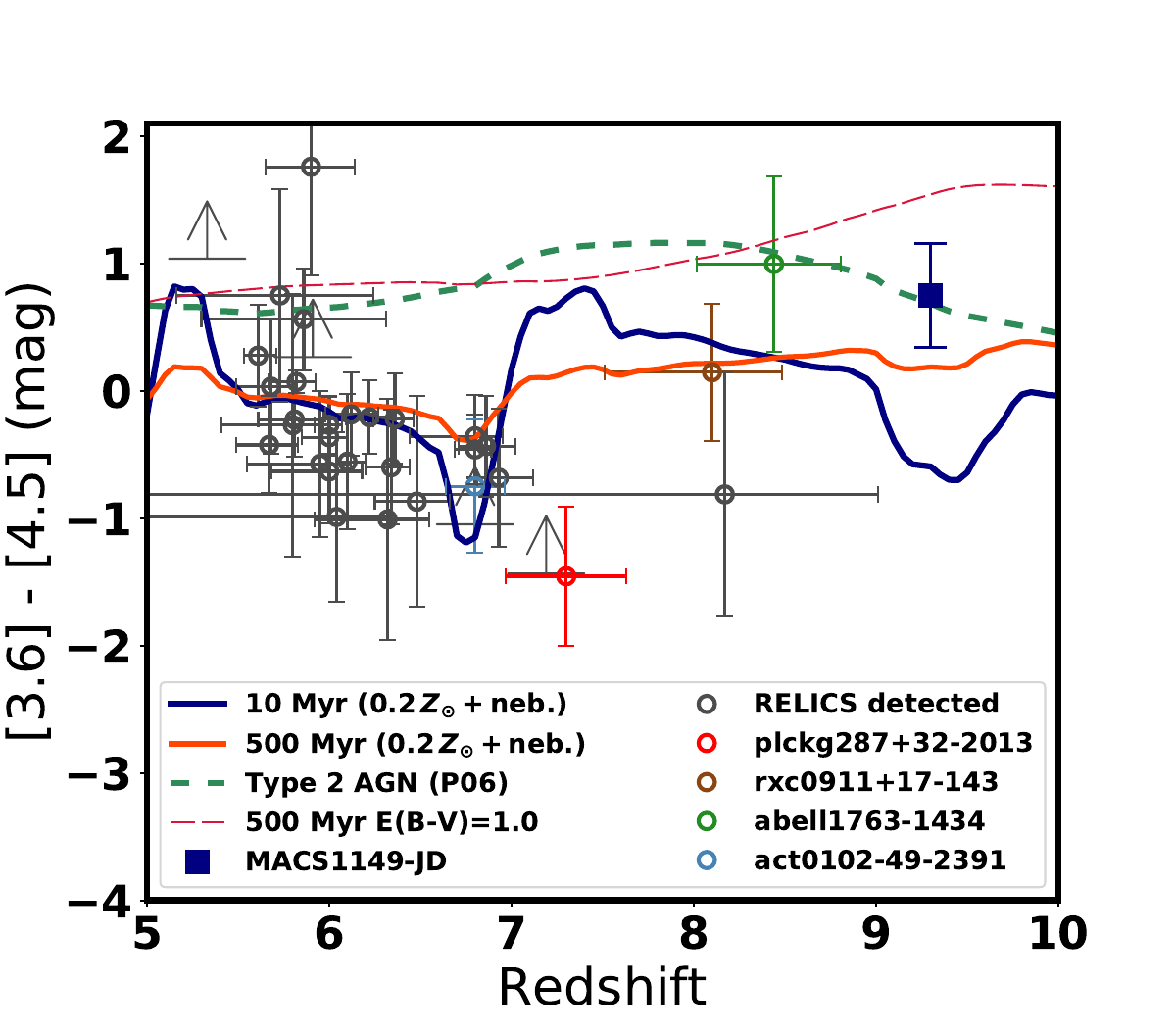}
    \caption{IRAC [3.6]$-$[4.5] color vs.~redshift for the galaxies in our sample with at least one IRAC S/N $>3$ detection (open circles with error bars). In red, brown, green, and blue open circles are PLCKG287+32-2013, RXC0911+14-143, Abell1763-1434, and ACT0102-49-2391, respectively, the objects we highlight later in \S\ref{results}. The dark blue filled square is MACS1149-JD from \cite{Hashimoto2018}, an object with evidence for an evolved stellar population at $z = 9.11$. Lines  are  tracks  from various models from BC03, showing the predicted colors for that model. Redshifts are calculated independently from these models (see Section \ref{sedfit}). 
    PLCKG287+32-2013 and ACT0102-49-2391 have colors consistent with the $z=6.6-6.9$ color bump due to [OIII]+H-$\beta$ emission in [3.6]. 
    Abell1763-1434 aligns well with older, dustier and AGN models.    
    }
    \label{fig:irac_colors}
\end{figure}

\subsection{Properties of the Sample}\label{stellarprops}
In this Section, we will first discuss the physical properties of the sample as a whole, presenting distributions of stellar mass, SFR, and age, and then later individual galaxies. Unless otherwise stated, we have used Method A for estimating the stellar properties presented here.

The quantity $\rm{L_{UV}/L^*_{UV}}$, the intrinsic luminosity of a source normalized by the characteristic luminosity for its redshift, informs the intrinsic brightness of our sample relative to that of the broader galaxy population. We calculate this by using $\rm{M_{UV} = H160} + 2.5\times\rm{log}_{10}(1+z) - 5\times \rm{log}_{10}(d_{pc}) + 5$, where H160 is the observed F160W flux in AB magnitudes (chosen because all objects are detected significantly in this band, and K-corrections using this band are negligible), and $d_{pc}$ is distance of the object in parsecs. We assume characteristic magnitude of $\rm{M^{*}_{UV}} = (-20.95\pm0.10) +(0.01\pm0.06)\times(z_{\rm{peak}}-6)$ from \cite{bouwens2015b}. We then correct for magnification in $\rm{M_{UV}}$ and convert both $\rm{M_{UV}}$ and $\rm{M^*_{UV}}$ to luminosity. In Figure \ref{fig:lstar} we present the distribution of $L_{UV}/L^*_{UV}$ for our sample, splitting the sample into those twice detected by IRAC, at least once detected, and the entire sample. $\sim95\%$ of our sample falls at or below $L_{UV}/L^*_{UV}$, with a few objects $\sim2L_{UV}/L^*_{UV}$, suggesting that we are probing a combination of galaxies characteristic for their redshifts as well as bright, perhaps unusual ones. The galaxies which are detected in both IRAC channels are, unsurprisingly, a larger percentage of intrinsically bright galaxies (the median value of $L_{UV}/L^*_{UV}$ for galaxies detected in both IRAC channels is 1.4, or -23.2 mag), however we do detect several $<0.5L_{UV}/L^*_{UV}$ galaxies. 

In Figure \ref{fig:distributions}, we show distributions of best-fit age of formation ($t_{\rm{form}}$), stellar mass and SFR, using assumptions from Method A in Section \ref{sedfit}. In each subplot, we again distinguish between the galaxies that are detected twice in IRAC, at least once in IRAC, and the entire sample. 
It is clear that galaxies in our sample with IRAC detections 
tend to take on a higher stellar mass, SFR, and formation age, possibly pointing to an observational bias. These are the same galaxies which tend towards intrisically brighter ($L_{UV}/L^*_{UV}>1$) in Figure \ref{fig:lstar}. We note that for a constant SFH, which we assume in our Method A SED fitting method, it is expected for at least some of these properties to be intrinsically correlated, as it will take a longer amount of time for a galaxy to build up a certain amount of mass at a given constant SFR. Notably, the overall distributions for the sample (and those of individual objects) between Methods A and B do not change appreciably, with the exception of formation age. 

As mentioned above, a possible effect at play in the sample distributions is an observational one: the galaxies with the brightest IRAC fluxes are the ones with detections, which will either be old galaxies with a strong Balmer break or young galaxies with strong emission lines. This is reflected in the age of formation distributions of each object in Figure \ref{fig:histograms}. Within the sample of high-$z$ candidates that received $\sim$30 hours of \emph{Spitzer} imaging, there are several examples of upper limits in flux constraining young galaxies (i.e., reaching depths of $\sim27$ mag without seeing a detection and thus revealing a very blue spectral shape and pointing to a young stellar population absent of dust). However, a large majority of the time (because the majority of our sample contains relatively shallow, $\sim5$ hour IRAC data), the comparison between detected and undetected galaxies is not necessarily a fair one. There could be galaxies with bright IRAC fluxes in clusters with shallow data that we are missing. Hence, the distributions in Figure \ref{fig:distributions} show a combination of the variety of stellar populations we are probing with our sample and our observational limits. 

In Figure \ref{fig:irac_colors}, we show the \textit{Spitzer}/IRAC [3.6]-[4.5] color as a function of redshift for a variety of models, including a young (10Myr) and dust-free template, and an evolved (500Myr) dust-free template, as well as an evolved template with E(B-V)=1.00. For comparison, we also include a Type 2 AGN model. These tracks were created from our set of BC03 + nebular continuum and emission templates described in Section \ref{sedfit} (Method A). Plotted over these models, we show the objects in our data which have at least one IRAC detection of S/N$>3$, highlighting select objects discussed below. Generally, we find that the IRAC colors in our sample trace the 10 Myr and 500 Myr models without dust, with some notable exceptions which we discuss in Sections \ref{emissionlinessect} and \ref{maxagesect}. 

\begin{figure*}[h!!!]
    \begin{subfigure}
        \centering
        \includegraphics[width=7.5cm]{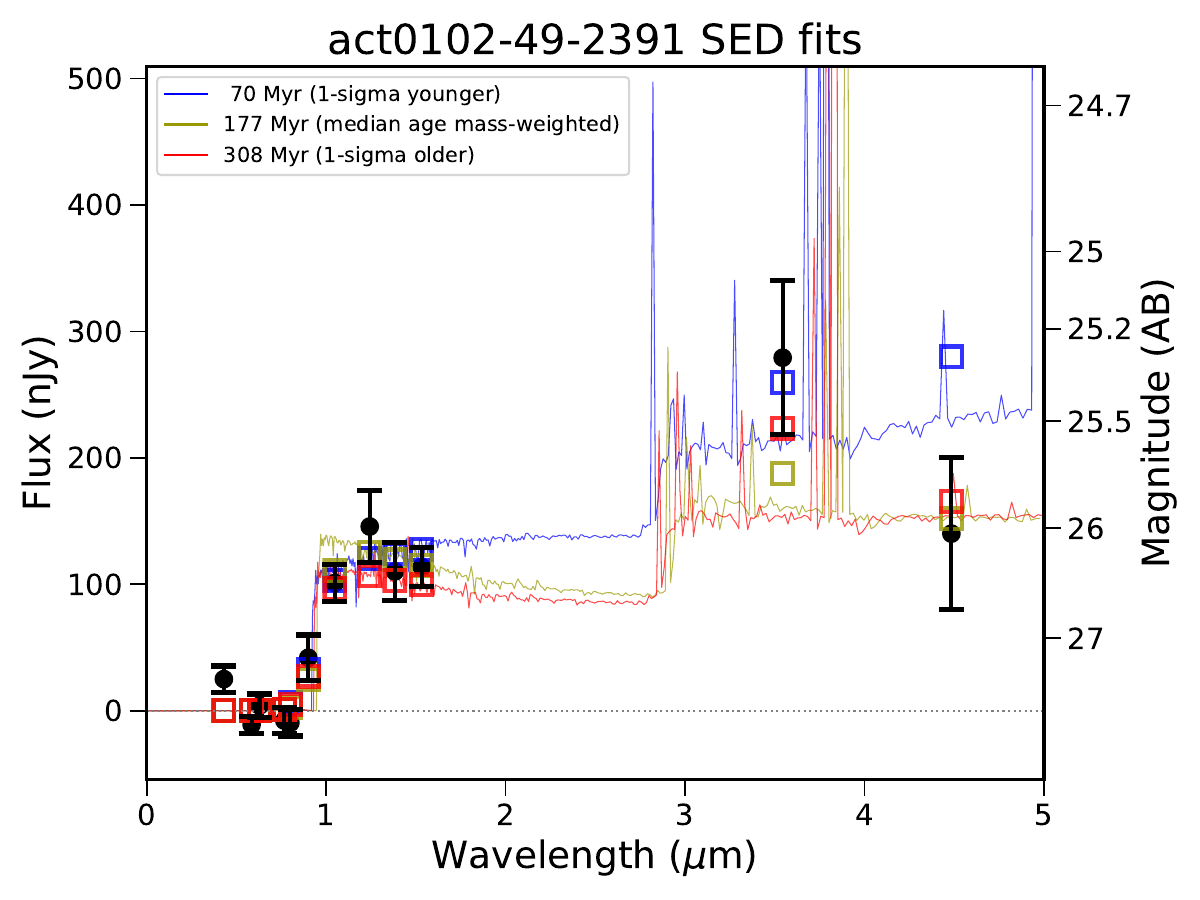}
    \end{subfigure}
    \begin{subfigure}
        \centering
        \includegraphics[width=9.5cm]{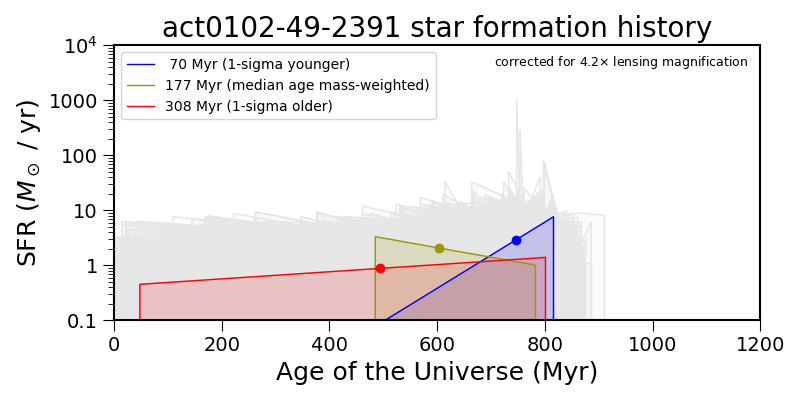}
    \end{subfigure}
    \begin{subfigure}
        \centering
        \includegraphics[width=7.5cm]{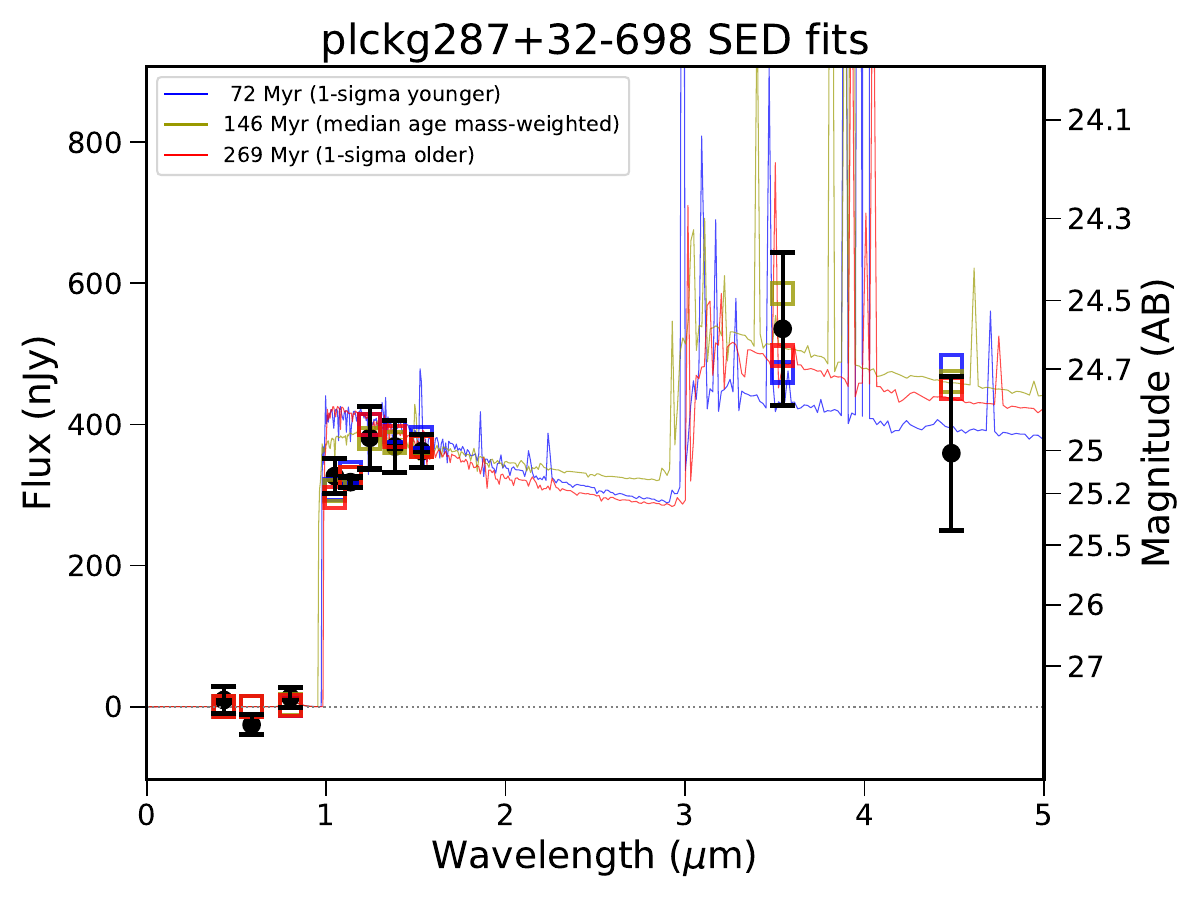}
    \end{subfigure}
    \begin{subfigure}
        \centering
        \includegraphics[width=9.5cm]{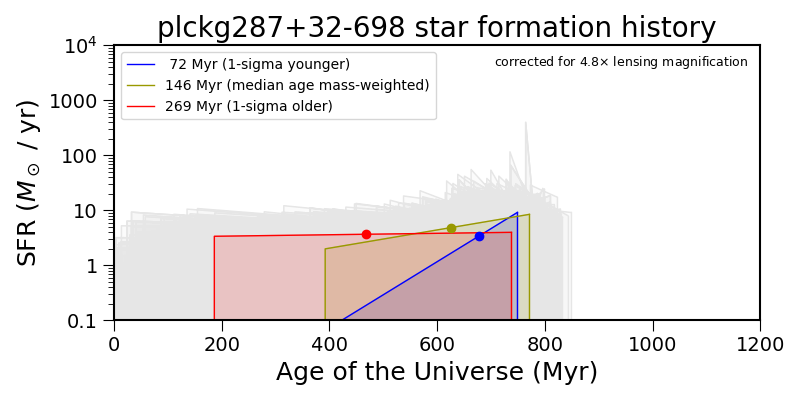}
    \end{subfigure}
    \begin{subfigure}
        \centering
        \includegraphics[width=7.5cm]{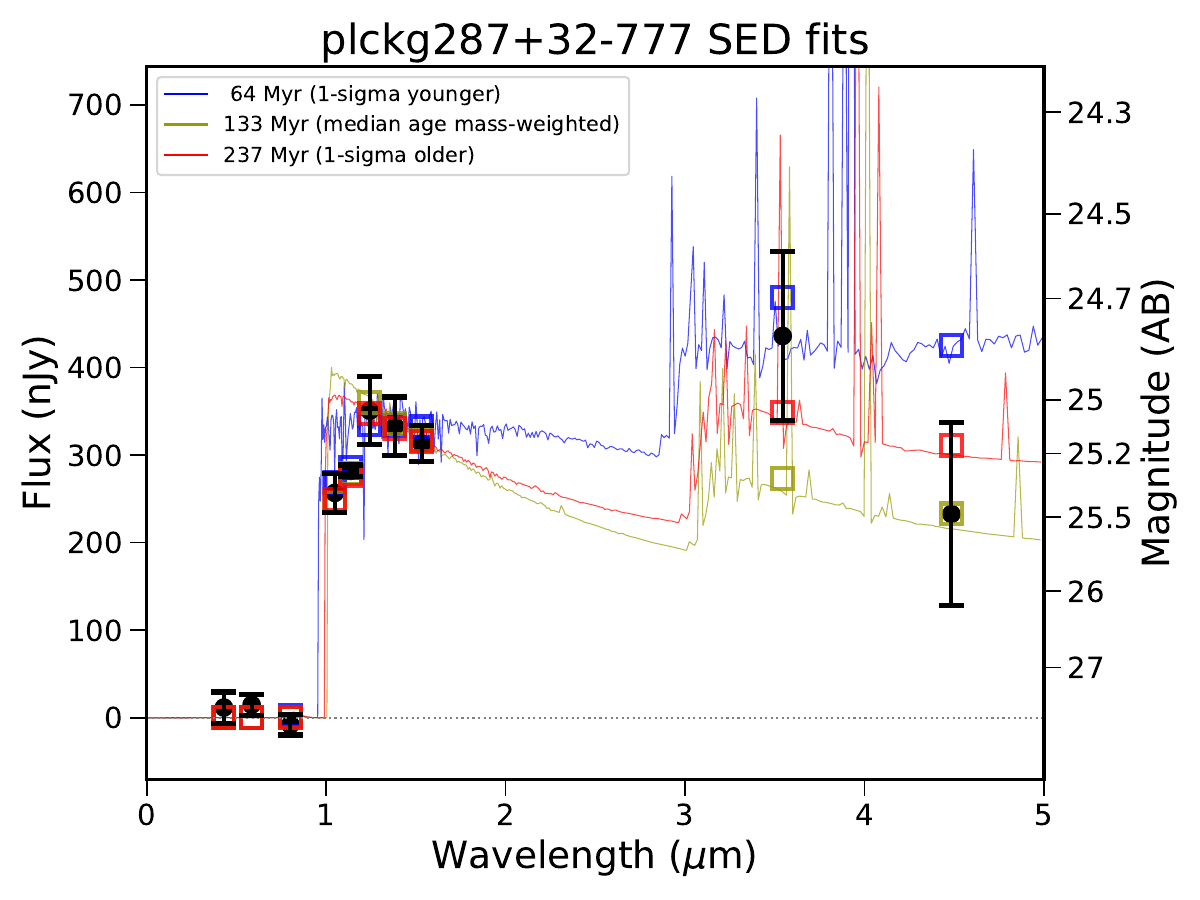}
    \end{subfigure}
    \begin{subfigure}
        \centering
        \includegraphics[width=9.5cm]{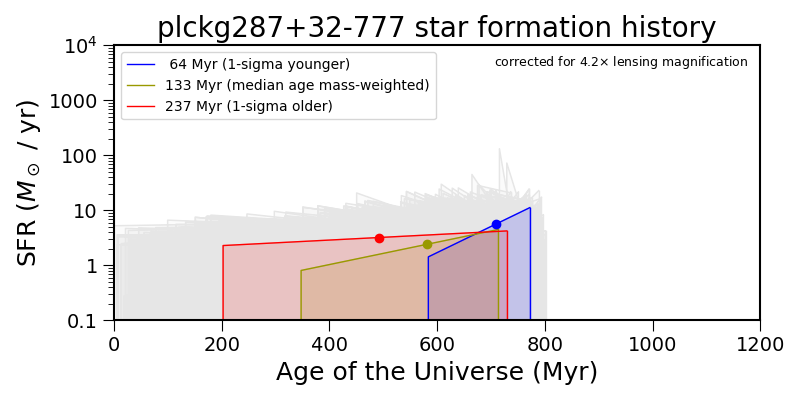}
    \end{subfigure}
    \begin{subfigure}
        \centering
        \includegraphics[width=7.5cm]{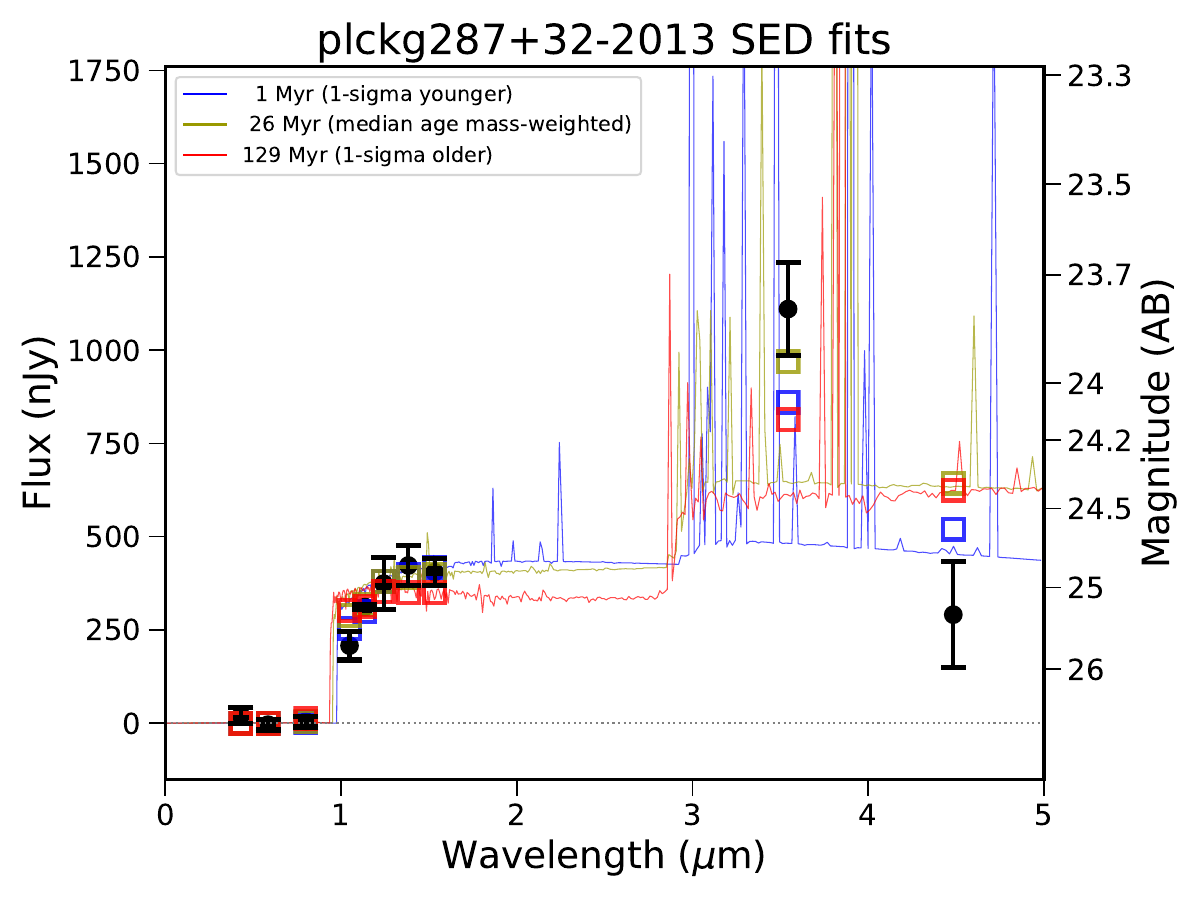}
    \end{subfigure}
    \begin{subfigure}
        \centering
        \includegraphics[width=9.5cm]{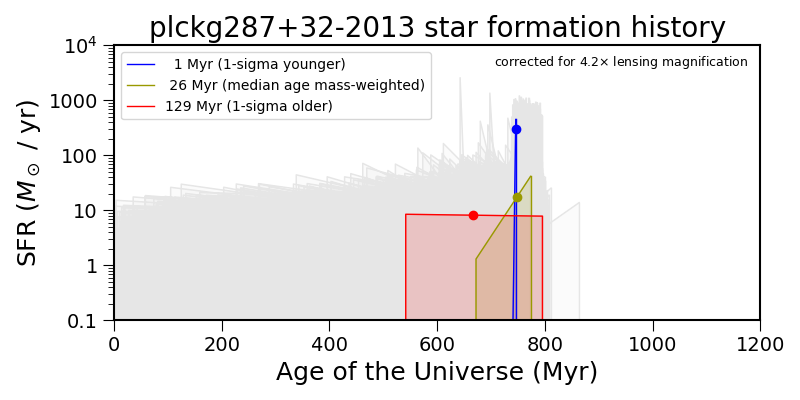}
    \end{subfigure}

\end{figure*}

\begin{figure*}
    \begin{subfigure}
        \centering
        \includegraphics[width=7.5cm]{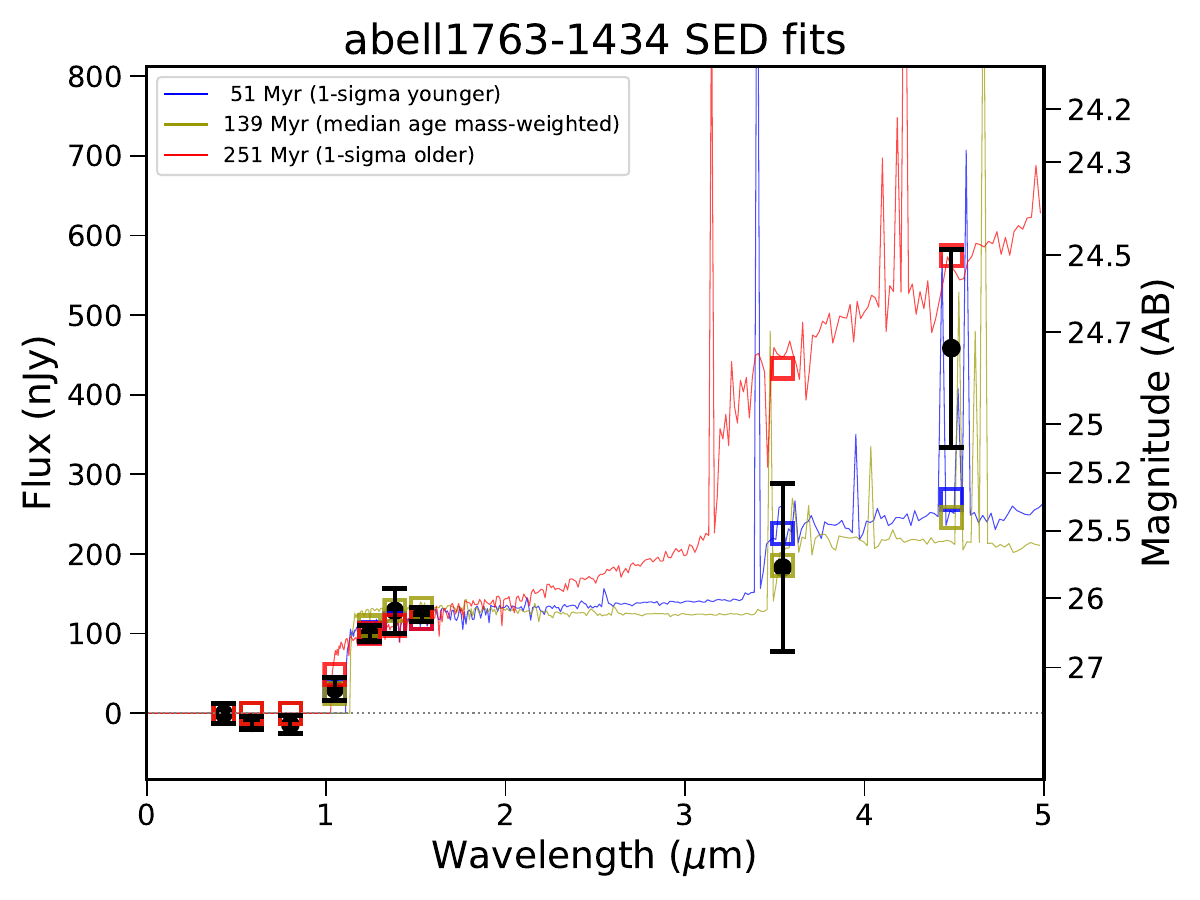}
    \end{subfigure}
    \begin{subfigure}
        \centering
        \includegraphics[width=9.5cm]{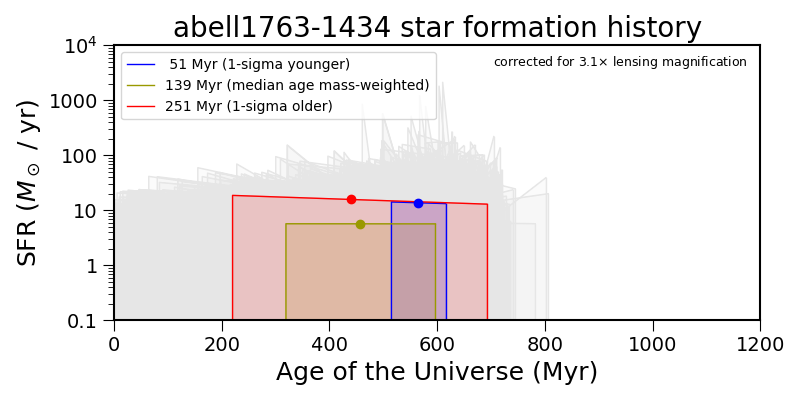}
    \end{subfigure}
        \begin{subfigure}
        \centering
        \includegraphics[width=7.5cm]{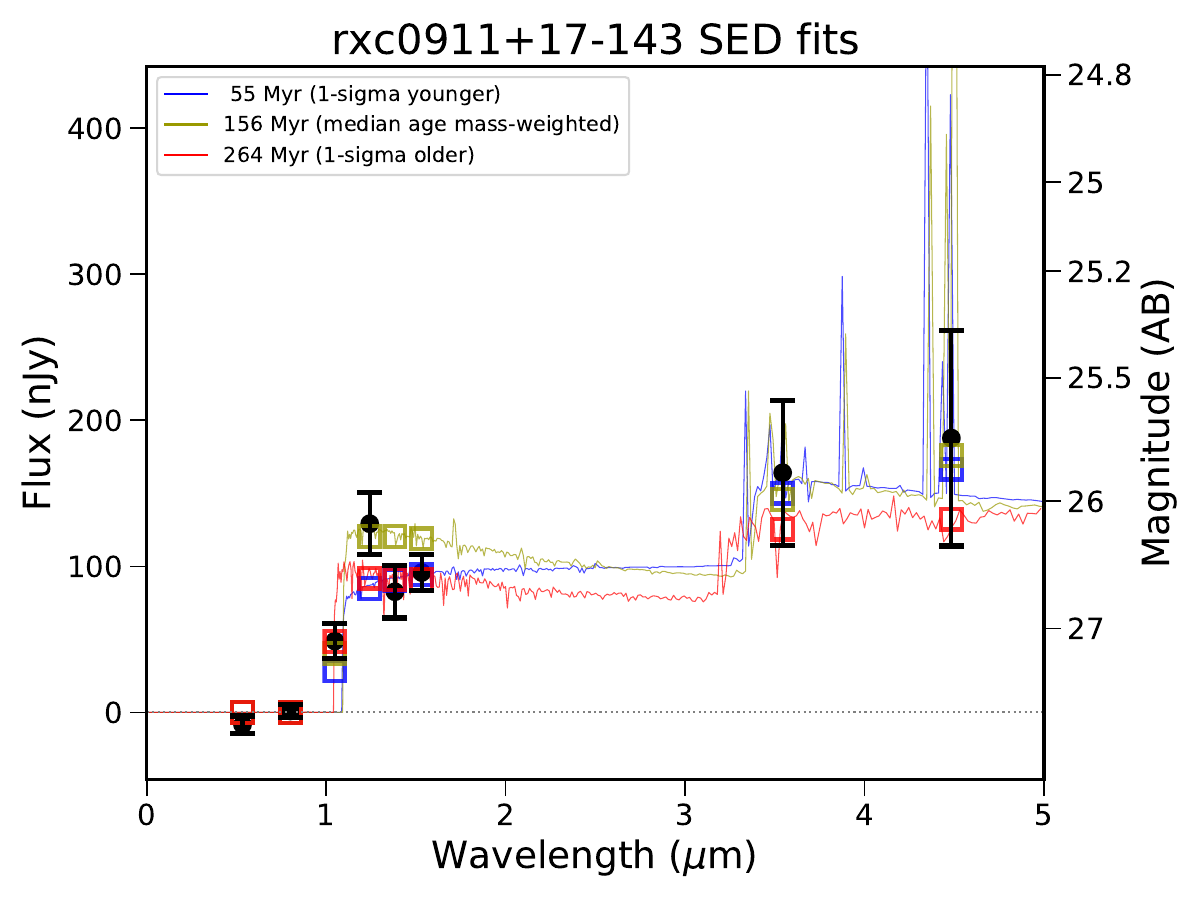}
    \end{subfigure}
    \begin{subfigure}
        \centering
        \includegraphics[width=9.5cm]{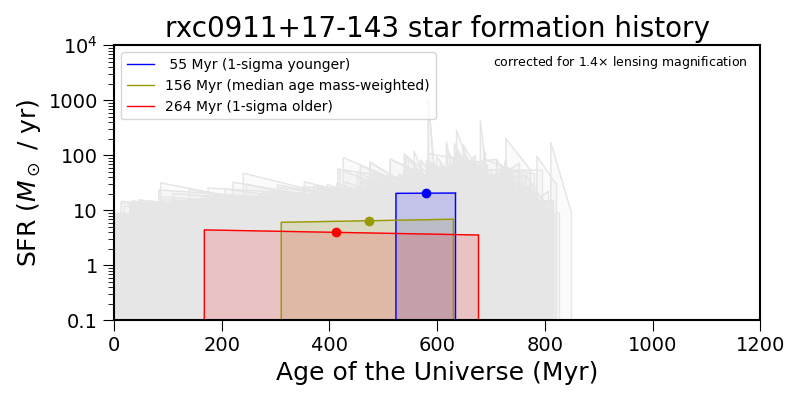}
    \end{subfigure}
    \caption{\textbf{Left: }Observed photometry (black), model photometry (open squares), and SEDs of the objects discussed in Sections \ref{emissionlinessect} and \ref{maxagesect} from Method B using BAGPIPES. Yellow template is an SED with the median mass-weighted age, blue is 1-$\sigma$ younger, and red is 1-$\sigma$ older. See Section \ref{biasandunc} for distinction between formation age and mass-weighted age. \textbf{Right: }SFHs of the templates on the left, in corresponding colors. Circles denote times of mass-weighted ages. In gray are all SFH realizations. }

    \label{fig:bagpipes}
\end{figure*}

\subsection{Detected Objects $z\geq6.5$}
RELICS was designed to find and characterize galaxy populations at high-redshift, apparently bright galaxies in a variety of fields that would be particularly good candidates for spectroscopic follow-up. For this reason, we highlight the objects we find to have the most informative data and be best candidates for follow-up. In Figure \ref{fig:bestfitseds}, we show the best-fit Method A SEDs for the 11 galaxies in our sample that have at least one $S/N>3$ IRAC detection and a best-fit redshift of $z\geq6.5$.  In blue, we show the best-fit high redshift ($z>4$) template and in red, the best-fit low redshift template ($z<4$) using the Method A assumptions described in Section \ref{sedfit}. The posteriors of stellar mass, formation age, star formation rate, specific star formation rate, and formation time for individual objects from Method A are shown in Figures \ref{fig:histograms}. In yellow/tan filled histograms, we show the results from Method A (BC03 templates, constant SFH), and in open dark green histograms, we show the results of Method B (BPASS, rising/declining SFH defined by Equation \ref{eqn:SFH}, extra birth cloud dust).

Of note in this sample is that several of the objects have likely redshifts around $z\sim6.6-7$, because of an elevated [3.6] magnitude, which indicates a strong [OIII]+H-$\beta$ emission line falling in ch1 at those redshifts. Additionally, we also see several objects which prefer a nearly  maximally old formation age; that is, the galaxies would have started forming $<100$ Myr after the Big Bang. 

\subsubsection{Objects With Potentially Strong Emission Lines}\label{emissionlinessect}
Of the objects highlighted in this work, the ones with blue IRAC [3.6]-[4.5] colors at $z=6.8-7.0$ include ACT0102-49-2391, Abell1758-1942, PLCKG287+32-698, PLCKG287+32-777, and PLCKG287+32-2013. Of these, we focus on three objects in PLCKG287+32: 2013, 698, and 777, as well as ACT0102-40-2391.

We show our analysis of these objects with BAGPIPES in Figure \ref{fig:bagpipes}: SEDs fit to the median and $\pm1\sigma$ in mass-weighted age template (note that we are now using mass-weighted age, defined in Equation \ref{eqn:mwage} instead of formation age as it is more easily constrained), and the median $\pm1\sigma$ in mass-weighted age SFHs in corresponding colors, as well as all allowed SFHs in gray. In PLCKG287+32-2013, one can see that the youngest (blue line) template shows a very recent, steeply rising burst of star formation. With a lower S/N detection, the range of solutions and SFHs for ACT0102-49-2391 are much dramatic.  

Additionally, we highlight these objects in Figure \ref{fig:irac_colors}. Of the four galaxies, PLCKG287+32-2013 has the highest S/N [3.6] detection ($\sim9\sigma$). 
EA$z$Y yields $z = 7.3 \pm 0.3$, while BAGPIPES finds $z = 6.8^{+0.2}_{-0.1}$.
As can be seen in Figure \ref{fig:irac_colors} (red and light blue circles), both PLCKG287+32-2013 and ACT0102-49-2391 show more extreme (blue) IRAC colors than most of the sample. In combination with the fact that they are consistent with the expected color decrease at $z\sim6.6-6.9$ for young, star-forming galaxies, we believe these are potentially strong [OIII]+H-$\beta$ emitters. The PLCKG287+32-2013 IRAC color corresponds to rest-frame equivalent widths of $\sim1000$\AA.\ Compared to galaxies observed at $z=0.2-0.9$ where [OIII]+H-$\beta$  equivalent widths are commonly observed to be $<100$\AA\, and extreme emission line galaxies (EELGs) are defined to be anything above that \citep{Mainali2020,Amorin2014}, this is a relatively high EW, and in line with the extreme emitters discovered by \cite{Endsley2020} (1000-4000\AA\ at $z\sim7$) and those discovered by \cite{Mainali2020} (500-2000\AA\ at $z\sim2-3$). These high EWs suggest a high value of the ionization parameter log(U), which corresponds to a hard spectrum and an ISM or even circumgalactic medium (CGM) replete with energetic ionizing photons. One possible consequence of this suggestion is the presence of an ionized bubble carved out by star formation (see, e.g., \citealp{Tilvi2020}) as a result of an ongoing burst of star formation (consistent with the youngest SFH for PLCKG287+32-2013 in Figure \ref{fig:bagpipes}). \cite{Endsley2020} argue that while galaxies with extreme [OIII]+H-$\beta$ may be experiencing a short-lived burst of star formation that, based on similarities to Lyman-continuum leakers (galaxies that ``leak" Lyman-continuum photons) at $z\sim3$, these objects tend to have very high escape fractions at least for the duration of the burst. Additionally, since these galaxies seem to make up $\sim20$\% of the $z\sim7$ population (though with large uncertainties), together they might be among the most effective objects at ionizing the IGM\footnote{Spectroscopic surveys can be used to target their Ly$\alpha$.}. 

Two of these objects (PLCKG287+32-698 and 777) are multiple images in a quadruply imaged system, first discovered by \cite{Zitrin2017}. 
In addition to finding the same redshift for both images analyzed here, we find convincingly similar physical properties with both Methods, as can be seen in Figure \ref{fig:histograms} ($\rm{M_*}\sim10^{8-9}M_\odot$, sSFR $\sim 10 \rm{Gyr}^{-1}$).
We note the third lensed image in this system, PLCKG287+32-2235, is included in our catalog but is not significantly detected in \textit{Spitzer}.
The fourth lensed image PLCKG287+32-2977 was discovered in WFC3/IR F110W imaging but lands outside RELICS imaging in other WFC3/IR filters and was thus excluded from the \cite{Salmon2018} selection and this work.

The last object we focus on in this section, PLCKG287+32-2013, yields some similarities to two galaxies confirmed by ALMA with [CII] by \citep{Smit2018}. They are of similar observed brightness (apparent magnitude $\sim24.9$), similar SFRs (within the uncertainties), and have similar IRAC colors. 
\cite{Smit2018} argue that an elevated [OIII]+H-$\beta$ EW suggests a higher [CII] EW, potentially meaning a more likely detection with ALMA follow-up for these objects. This would allow us to place them on the infrared excess-UV slope (IRX-$\beta$) relation. Additionally, such extreme EWs are indicative of potential Ly-$\alpha$ emission \citep{Roberts-Borsani2016,Oesch2015,Zitrin2015,Stark2016}, and CIII] (e.g., \citealp{Hutchison2019}) presenting opportunities for ground-based infrared spectroscopy. 
\subsubsection{Gemini Observations of PLCKG287+32-777}
We take this opportunity to note that PLCKG287+32-777 was spectroscopically observed with Gemini South (GS-2018A-Q-901; PI: Zitrin), and we publish here the first results from these observations. It was first observed for 2 hrs with Flamingos-2 (R3000 Grism + J-band filter). A 4-pixel ($0.72\arcsec$) wide longslit was used, leading to an average resolving power of $R\sim1300$, with a nod size of  $\pm1.5\arcsec$. These observations were designed to detect potential CIV emission, although the detection limit is quite high (e.g., a $3\sigma$ limit per spectral pixel of $\simeq1-1.5\times 10^{-16} \rm{erg/cm}^2/$\AA\, for a spectral line width of 100-300 km/s).  

The object was also observed with Gemini Multi-Object Spectrograph (GMOS) in longslit Nod \& Shuffle mode, with a 1" wide slit and a $\pm2.5\arcsec$ nod size. Observations took place using the R400 grism + z-band filter, and two different central wavelengths (760 nm, 795 nm), for $\simeq 4.3$ hrs. These observations were aimed to detect Ly-$\alpha$, and have a nominal $3\sigma$ depth per spectral pixel of  $4.5\times10^{-17} \rm{erg/cm}^2/$\AA\, for a 500 km/s Ly-$\alpha$ line at $z=6.8$. At higher redshifts, the sensitivity drops, reaching half the SNR for Ly-$\alpha$ at $z\simeq7.3$.

Both the F-2 and GMOS longslits also covered a second image of the lensed galaxy (2977), however the nod size in the GMOS observations only allowed for a credible search around the first object. No prominent emission line was detected in either observation for either image, suggesting that Ly-$\alpha$ may not be easily escaping this galaxy and the presence of extremely strong CIV emission is likely ruled out. However, we note that the data merit a more careful inspection, and we leave a more quantitative examination of these data for future work. 

\subsubsection{Objects Preferring Maximally Old Formation Age}\label{maxagesect}
In addition to galaxies with potentially strong nebular emission, there are a subset of galaxies in our sample that show a strong preference with Method A and some preference with Method B, for a maximally old solution, with the earliest possible formation time, including ACT0102-49-2551, RXC0911+17-143, MACS0553-33-219, Abell1763-1434, and CL0152-13-191. Here we focus on Abell1763-1434 and RXC0911+17-143, both of which were initially introduced by \cite{Strait2020a}. The former has since received deeper data ($\sim$30 hours, from $\sim$5 hours in \citealp{Strait2020a}).  

As has been noted previously in this work, formation age is a difficult parameter to constrain, since the beginning of SFHs tend to leave little imprint on a photometric SED. The results tend to be reliant on the implicit prior of the assumed SFH, which in the case of our Method A is constant. In Method B, we report mass-weighted age instead, since this quantity can be more reliably fit for. We will discuss both here. 

In the fifth panel of Figure \ref{fig:bagpipes}, we show our analysis with BAGPIPES of Abell1763-1434, where we plot again median $\pm1\sigma$ mass-weighted age templates. In all cases, these templates represent relatively evolved stellar populations, reflected by the SFHs on the right panel which ``turn on" relatively early. The oldest solution prefers a gently declining SFH, while the others are relatively constant. There are some solutions that prefer a high [OIII]+H-$\beta$ EW (i.e., a young, high sSFR template rather than an evolved one with a Balmer break), which falls in IRAC ch2 at the redshift of this galaxy ($z\sim8.4$), but overwhelmingly, an evolved stellar population is preferred. 

We see a similar case in RXC0911+17-143, shown on the bottom panel of Figure \ref{fig:bagpipes}: this time with both IRAC fluxes elevated, there is evidence of a Balmer break, although the detections are lower S/N compared to Abell1763. The SFHs reflect a similar preference for a declining or constant, but relatively stagnant SFH for this object. When these objects are compared to the rest of our sample in Figure \ref{fig:irac_colors}, both are consistent with an evolved and/or dusty solution. The red IRAC colors make these objects stand out from the rest of our sample as being closer in color-redshift space to MACS1149-JD, a spectroscopically confirmed galaxy with evidence for an early formation time \citep{Hashimoto2018}.

Similarly, ACT0102-49-2551, CL0152-13-191, and MACS0553-33-219 show evidence of a Balmer break. In the cases of these galaxies, both IRAC fluxes are elevated, making it less likely that the elevated flux is due to emission lines alone. The EW of [OIII]+H$\beta$ EW for MACS055-33-219 and ACT0102-49-2551 would need to be $\sim1000$\AA\ and $\sim 300$\AA, respectively. In the case of CL0152-13-191, the \textit{Spitzer} fluxes are elevated by $\sim2.3$ mags, and we do not find any well-fit young solutions in either method. Further followup on these galaxies will be necessary for proper modeling.

As argued by \cite{Strait2020a} and \cite{Roberts-Borsani2020}, IRAC excess in [4.5], while often attributed to high values of [OIII]+H-$\beta$ (which fall in ch2 at $z>7$), can be equally (and sometimes more favorably) explained by a strong Balmer break, suggesting an evolved stellar population and/or dust. \cite{Roberts-Borsani2020} (and previously \citealp{Hashimoto2018}) goes on to suggest ways of distinguishing these solutions, including the use of the ALMA to detect the [OIII] 88$\mu$m line, whose strength is related to the [OIII] 5007\AA\ line. We suspect that the galaxies discussed in this section likely have Balmer breaks and evolved stellar populations.

\subsection{SFHs and Age Constraints}
 In addition to the Method B BAGPIPES run described in Section \ref{sedfit}, we also performed an alternative run for the 11 galaxies highlighted in this paper, in order to explore a wider parameter space and test the effect of our assumptions on the difference in results between Methods A and B. For this run, we allowed redshifts to range from $z=0-12$, adopted an SMC dust law, and allowed for a freely-varying constant SFH and a recent burst (as in, e.g., \citealp{Hashimoto2018,Roberts-Borsani2020}). The effect that these changes had on the candidates described here was mainly due to the newly allowed burst in the SFHs. In some cases, this increased the median age and in some cases, it lowered the median age. Our conclusion from this test is that our data are unable to meaningfully constrain the SFHs of this sample. This is also evident in Figure \ref{fig:bagpipes}, where it is clear that there are many SFH solutions that fit the data. We plan to explore what data would be necessary to constrain SFHs meaningfully, including testing \textit{JWST} and ALMA data, in a future work. 

\subsection{Demoted Objects}
Of the 207 galaxies for which we could extract reliable \emph{Spitzer} fluxes, 23 have non-trivial peaks in redshift at $z<2$, revealing a $\sim10$\% contamination rate for \textit{HST}-selected Lyman-break galaxies.
Notably, the bright $z\sim6$ galaxies behind RXS0603+42 (both north and south HST pointings) had a high ``failure" rate. Of the 13 high-redshift candidates behind this cluster, we were able to extract \emph{Spitzer} fluxes for 8. Of those, only 2 remain likely at high redshift. The rest are likely $z\sim1$ galaxies or brown dwarfs.

%
%

\section{Future Data}\label{futuredata}
Ultimately, spectroscopic followup will be necessary to place strong constraints on properties of these galaxies, such as dust content, metal enrichment, and ionization field. There are a number of existing and future telescopes that will be well-suited to this task:

Ground-based infrared spectroscopy with telescopes such as Keck will allow for a search of rest-frame UV emission lines such as Lyman-$\alpha$ and CIII]. There are a number of studies detailing how such observations can aid in constraining SFR, metallicities, AGN activity, and ionization with these emission lines (e.g., \citealp{Finkelstein2013,Stark2016,Nakajima2018,Hutchison2019,LeFevre2019}). As discussed above, the presence of strong [OIII]+H-$\beta$ emission makes detection of Lyman-$\alpha$ and CIII] more likely.

A millimeter/sub-millimeter telescope such as ALMA could be used to determine the presence of [OIII]$88\mu$m or [CII]$158\mu$m, which would lead to insight on the dust and metal content (e.g., \citealp{Roberts-Borsani2020,Bradac2017}), constraints on the strength of [OIII]5007\AA\ \citep{Hashimoto2018,Roberts-Borsani2020}, and even kinematics (e.g., \citealp{Smit2018}). Placing high-redshift galaxies on the IRX-$\beta$ relation is valuable for understanding typical dust properties relative to the shape of the UV continuum. We calculate the UV slopes\footnote{We calculated beta slopes from the photometry, excluding F105W for PLCKG287+32-2013, which has the potential to for contamination from Lyman-$\alpha$ absorption (if included, the $\beta$ slope for PLCKG287+32-2013 is $-0.4\pm1.2$).} of PLCKG287+32-2013 and ACT0102-49-2391 from the photometry to be $-1.7\pm1.1$ and $-1.7\pm1.3$, respectively. For $\beta$ slopes of these values, following the \cite{Meurer1999} relation for local galaxies, one would expect values of IRX (log($L_{IR}/L_{UV}$)) of around $\sim0-1$. However, there has been high scatter observed in this relation, and in fact the $z>5$ galaxies studied by, e.g.,  \cite{Willott2015,Capak2015,Knudsen2017,Smit2018,Fudamoto2020} are found to have lower-than-expected IRX. This decrement is still present even when assuming a steep attenuation law such as SMC, perhaps explained by a higher dust temperature at high redshift.

With capabilities out to 28 microns, \textit{JWST} will revolutionize high-redshift galaxy spectroscopy and imaging and allow us to do detailed analyses of the rest-frame UV and optical spectrum for galaxies in the $z\gtrsim6$ regime. Notably, spectroscopic (and inferred photometric) measurements of [OIII]+H-$\beta$ and their strength relative to other rest-frame UV and rest-frame optical emission lines will aid in our understanding of ionization field, ionizing photon production, oxygen abundance and sSFR.

\section{Conclusions}\label{concls}
We present an analysis of new \textit{Spitzer}/IRAC imaging for the high-$z$ candidates in the RELICS survey, providing a full \textit{HST} and \textit{Spitzer} catalog, with galaxy properties, redshifts, and magnifications for 207 galaxies likely at $z\geq5.5$. We present the distributions of stellar properties of the sample using templates from BC03 and BPASS + nebular emission, and highlight 11 galaxies that have the highest redshifts ($z\geq6.5$) and at least one S/N detection $>3$ in IRAC. We go into further detail for six of those objects. Our main conclusions are as follows:

\begin{itemize}
    \item While $\sim95\%$ of our sample are characteristic for their redshift ($L_{UV}/L^*_{UV}<1$), there are a few that are intrinsically brighter ($\sim2L_{UV}/L^*_{UV}$). Within our sample, we see a variety of stellar populations, from very small at $2.1\times10^5\rm{M_{*}}$ to very massive at $4.2\times10^9\rm{M_{*}}$, and from very young (forming $>800\rm{Myr}$ after the Big Bang) to very old (forming $<100\rm{Myr}$ after the Big Bang).
    \item Along with this paper, we are releasing our full \textit{HST} + \textit{Spitzer}/IRAC photometric catalog\footnote{\href{http://victoriastrait.github.io/relics}{victoriastrait.github.io/relics}}, as well as results from both methods of SED fitting described in this work and magnifications from publicly available lens models. 
    \item We find that PLCKG287+32-2013, one of the brightest $z \sim 7$ candidates known (F160W mag 24.9), has strong evidence for strong [OIII]+H-$\beta$ emission (EW $\sim1000$\AA), as suggested by its elevated \textit{Spitzer} [3.6] flux. We believe that this galaxy is experiencing an ongoing burst of star formation. 
    \item We find a similar object, ACT0102-49-2391, which although less luminous, also reveals an elevated [3.6] flux, and falls around $z\sim6.6-6.9$, again suggesting strong nebular emission.
    \item We find two objects, Abell1763-1434 and RXC0911+17-143, that show evidence for an evolved stellar population, i.e., that they formed very early ($<100$ Myr after the Big Bang). We believe that these galaxies show good evidence of Balmer breaks, and their IRAC colors are consistent with evolved (500 Myr) or dusty (E(B-V)=1.00) galaxy models.
    \item While several of our $z\geq6.5$ galaxies which are detected in IRAC prefer a nearly maximally old solution, there may be other explanations for their bright \textit{Spitzer} fluxes, such as dust or extreme line emission. Disentangling these degeneracies will only be possible with, e.g., \textit{JWST}, Keck, Thirty Meter Telescope, and/or ALMA observations. 
    \item Through the exploration of SFHs, we find that formation age, which is commonly used as the measure of age in high-redshift galaxy studies, is difficult to constrain. We report it here for comparison with other works, but also present mass-weighted age, a more easily constrained property.
\end{itemize}

The galaxies presented in this work will be excellent targets for follow-up existing telescopes like Keck and ALMA, as well as with future telescopes such as \textit{James Webb}, Thirty Meter Telescope and Giant Magellan Telescope, as they are apparently bright but intrinsically faint, and likely the dominant galaxy population at these epochs.
\section*{Acknowledgements}
Based on observations made with the  {\it Spitzer} Space Telescope, which is operated by the
Jet Propulsion Laboratory, California Institute of Technology under a
contract with NASA. Also based on observations made with the NASA/ESA Hubble Space Telescope,
obtained at the Space Telescope Science Institute, which is operated
by the Association of Universities for Research in Astronomy, Inc.,
under NASA contract NAS 5-26555.
Support for this work was provided by NASA
through  ADAP grant 80NSSC18K0945, NSF grant AST 1815458,   NASA/HST grant {\it HST}-GO-14096, and through
an award issued by JPL/Caltech. VS also acknowledges support through Heising-Simons Foundation Grant \#2018-1140.

This work also refers to observations obtained at the international Gemini Observatory, a program of NSF’s OIR Lab, which is managed by the Association of Universities for Research in Astronomy (AURA) under a cooperative agreement with the National Science Foundation on behalf of the Gemini Observatory partnership: the National Science Foundation (United States), National Research Council (Canada), Agencia Nacional de Investigaci\'{o}n y Desarrollo (Chile), Ministerio de Ciencia, Tecnolog\'{i}a e Innovaci\'{o}n (Argentina), Minist\'{e}rio da Ci\^{e}ncia, Tecnologia, Inova\c{c}\~{o}es e Comunica\c{c}\~{o}es (Brazil), and Korea Astronomy and Space Science Institute (Republic of Korea). The guaranteed observations used here were obtained through Ben-Gurion University's (BGU; Israel) time on Gemini, following a MoU between BGU and Gemini/AURA.

\bibliographystyle{apj}
\bibliography{vstrait}

\end{document}